\documentclass[aps,prb,showpacs,raggedbottom,nobalancelastpage,amssymb]{revtex4}
\usepackage{amsmath}
\usepackage{amsfonts}
\usepackage{graphicx}
\usepackage{float}
\usepackage{appendix}
\usepackage{dsfont}
\usepackage{amssymb}
\usepackage{bm}
\usepackage{color}
\usepackage{ulem}
\usepackage{soul}
\usepackage{natbib}
\setstcolor{red}
\usepackage{nccmath}
\usepackage{array}

\newcommand{\beq}{\begin{equation}}
\newcommand{\eneq}{\end{equation}}
\newcommand{\be}{\begin{equation}}
\newcommand{\ee}{\end{equation}}
\newcommand{\bea}{\begin{eqnarray}}
\newcommand{\eea}{\end{eqnarray}}

\usepackage{multirow}
\usepackage{wasysym}

\begin{document}
\title{Tunable spin/charge Kondo effect at a double superconducting island connected to two spinless quantum wires}
\author{Domenico Giuliano$^{(1,2)}$,  Luca Lepori$^{(1,3)}$, and Andrea Nava$^{(1,4)}$ }
\affiliation{
$^{(1)}$ Dipartimento di Fisica, Universit\`a della Calabria Arcavacata di 
Rende I-87036, Cosenza, Italy \\
$^{(2)}$ I.N.F.N., Gruppo collegato di Cosenza, 
Arcavacata di Rende I-87036, Cosenza, Italy \\
$^{(3)}$ Istituto Italiano di Tecnologia, Graphene Labs, Via Morego 30, I-16163, Genova, Italy \\  
$^{(4)}$ International School for Advanced Studies (SISSA), Via Bonomea 265, I-34136 Trieste, Italy
 }
\date{\today}

\begin{abstract}

We propose  that a  pertinently engineered double superconducting island connected to 
two spinless one-dimensional conducting  leads  can work as a tunable (iso)spin Kondo and charge Kondo system,
 with  the lead index regarded as an effective isospin degree of freedom.
We evidence how, by tuning a single  gate voltage applied to the island, it is possible to 
make the system switch from the (iso)spin Kondo to the charge-Kondo phase,
passing across an intermediate   phase, in which the Kondo impurity is effectively 
irrelevant for the low-temperature behavior of the system. 
  Eventually, we evidence how to probe the various phases by measuring the ac conductance 
tensor of our system, by emphasizing  the features that should allow to identify the onset of the so far 
quite elusive ``charge-2''  charge-Kondo effect. 
 \end{abstract}

\maketitle

\section{Introduction}
\label{intro}

The Kondo effect has been experimentally  seen for the first time as an upturn in the 
resistivity of metals doped with magnetic impurities, as the temperature $T$ goes below  the 
nonuniversal Kondo temperature $T_K$, typically of the order of a few Kelvin, or less \cite{kondo_0,hewson}. On the theoretical 
side, the Kondo effect was readily explained in terms of a dynamical screening of the single-impurity 
magnetic moment by means of the spin density of itinerant electrons in the metal \cite{kondo_0,hewson}.
As $T \to 0$,  the impurity moment is fully screened, which allows for trading the impurity  for  a local (Kondo) spin singlet, acting as a 
scattering center that forbids electrons from accessing the impurity site \cite{Noz1974,hewson}. 
The Kondo spin singlet provides  one of the few theoretically well-understood examples of strongly correlated
states of matter. For this reason, soon after its discovery and its theoretical explanation, Kondo effect 
began to be used as a paradigmatic test bed for a number of remarkable analytical, as well as numerical, 
methods to study strongly correlated systems, including the well-celebrated Wilson's numerical renormalization 
group (RG)  technique \cite{wilson,bulla}. On top of that, it has been found how 
a peculiar realization of the effect, such as  the ``overscreened'' one, in which more than one 
itinerant electron ``channel'' contributes to screen the magnetic impurity, yields to a 
novel phase of matter, which, differently from the local spin singlet, 
cannot be described within the ``standard'' Landau's Fermi-liquid  framework \cite{afl_2}.

Recently, a renewed interest has arisen in the Kondo effect \cite{kouwenhoven01}, due to the possibility of 
realizing it in a controlled way in mesoscopic systems with tunable parameters, such as 
semiconducting  quantum dots with metallic leads \cite{alivisatos96,kouwenhoven98,gg98_1,gg98_2}, in which Kondo effect is expected to appear as 
an upturn of the conductance, rather than of the resistivity, across the dot connected to the leads,
or with  superconducting leads \cite{avish,choi,gbab},  in which Kondo effect should be evidenced by a change 
in the behavior of the subgap (Josephson) supercurrent across the dot, when the leads are held at a 
fixed phase difference $\varphi$. In addition, since the  effect is merely due to spin dynamics, 
it has been proposed that a ``spin-Kondo'' effect can take place in systems with itinerant, low-energy 
excitations carrying spin, but not charge, such as XXZ spin-1/2 chains \cite{affleck_eggert,furusaki98,grt}, which can be for instance 
realized by loading cold atoms on a pertinently designed optical lattice \cite{grst_13,gst}, or frustrated $J_1-J_2$
spin chains with $J_2 /J_1$ tuned at the ``critical'' value at the phase transition between 
the spin liquid- and the dimerized-phase of the system \cite{sorensen}. In fact, realizations 
of the spin-Kondo effect have recently been proposed at junctions of quantum spin chains \cite{crampe,tsv_1,tsv_2,gst_1}, or of 
one-dimensional arrays of Josephson junctions \cite{giuliano09,cmgs,Giuliano_2013}. Finally, it is    worth mentioning 
the possibility, that has been recently put forward, that a remarkable ``topological'' Kondo effect might 
arise, in which the impurity spin is realized by means of Majorana modes emerging at the interface between 
topological superconductors and normal conductors (or the empty space) \cite{beri,beri_cooper,altland_beri} and, more generally, 
the striking similarity between the Kondo physics and the hybridization between a Majorana mode and 
the itinerant electrons in a metal connected to the topological superconductor \cite{ga_w}.

A key point about Kondo effect is that, asite from all the spin dynamics underneath, in order for 
the effect to take place, one generally needs an impurity with a twofold-degenerate ground state, 
which is able to switch from one state to the other via quantum number exchange processes with 
itinerant particles from the medium into which the impurity is embedded. In this respect, 
a number of proposals have been put forward in which Kondo effect is  associated to 
{\it charge}, rather than to spin degeneracy in the impurity ground state. Such a charge-based 
version of Kondo effect is typically dubbed ``charge-Kondo" (CK) effect, to distinguish it 
from the ``standard'' ``spin-Kondo'' (SK) effect. The  CK effect was originally proposed as a 
possible mechanism, related to the ``negative-$U$'' Anderson model, able to induce a 
charge-dual version of the highly-correlated, heavy-fermion ground state \cite{cot}. 
The CK effect has later on been theoretically studied in dots connected to bulk leads \cite{matve_1},
in single-electron transistors \cite{zar,bux,lhp}, as well as in generalizations of the 
negative-$U$ Anderson model \cite{corn,dz,ander}, also involving optical lattice systems \cite{Laad_2009}.
 Over all, there are two main different realizations of CK effect. The former 
one does not require superconducting correlations. It  consists in realizing 
the two degenerate impurity states at a quantum point contact with a finite charging energy 
tuned, by means of an external gate voltage, nearby the degeneracy point between two 
states differing by a single electron charge $e$ (``charge-1'' CK). This proposal has been theoretically put forward 
in, e.g., Refs.[\onlinecite{matve_1,furusaki,zar,bux,lhp}] and eventually led to the experimental 
observation of the effect, discussed in Ref.[\onlinecite{expc1}]. At variance, a different realization of 
CK effect, in which the degenerate impurity states differ by a charge $2e$ (``charge-2'' CK), 
has been proposed in, e.g., Refs.[\onlinecite{cot,corn,dz,ander}]  (a detailed discussion about 
the two different realizations of CK is provided in, e.g., Ref.[\onlinecite{ion}]).

Notwithstanding the great interest in CK effect, witnessed by the large number of papers on 
the topic, a clear-cut experimental verification of charge-2 CK effect is still lacking (differently 
from what happens for SK and for charge-1 CK effect).  For this reason, in the last years there has been an increasing
interest in realizing CK effect in a controlled way, in systems with tunable parameters. For instance, 
it has been proposed to realize CK effect in mesoscopic superconductors coupled to 
normal metals \cite{ion}, in negative-$U$ quantum dots with superconducting electrodes 
\cite{negu}, and even in double quantum dot, in which the effect should be mediated by the 
Coulomb repulsion between the electrons at the double dot \cite{taba}. In general, defining an 
appropriate tunable device to probe CK effect, possibly in comparison with the more ``standard''
SK effect, is still an open challenge, also in view of the potential relevance of CK effect 
to explain the physics of, e.g., superconductivity in PbTe doped with Tl \cite{dz}, or of 
impurities formed at dots in ${\rm LaSrO_3/SrTiO_3}$ interfaces \cite{lao_sto}

\begin{figure}
 \center
\includegraphics*[width=0.5 \linewidth]{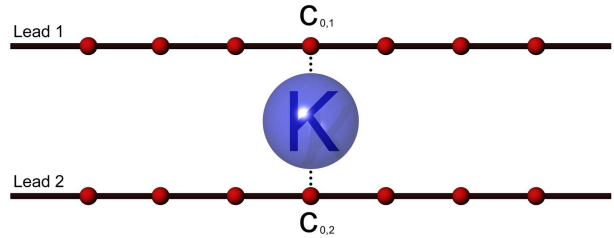}
\caption{ Sketch of our proposed system: the two quantum wires, 
are  represented as one-dimensional lattices, while the  tunable Kondo impurity is 
realized   by connecting the leads to a double superconducting island with pertinently 
chosen parameters (see the main text).} 
\label{ladder}
\end{figure}

In this paper, we propose to realize  charge-2 CK effect at a ``minimal'' tunable device in which
one may in principle switch from SK, to CK effect by just acting onto a limited 
number of system's parameters (ideally, one parameter only). Our system, which we 
sketch  in Fig.\ref{ladder} in its ``minimal'' version, consists of two spinless conducting 
fermionic channels (the ``leads''), connected to a ``tunable'' effective Kondo impurity 
$K$. In particular, we
propose to  realize the tunable  Kondo impurity by means of a pertinently designed
double superconducting island   hosting four Majorana modes emerging at the endpoints of two 
spinless quantum wires deposited on top of it (see Fig.\ref{topo_island} for a sketch of the setup).
Aside from technical details related to the design of our
system, it is based on by now well-established features concerning the interplay between emerging 
Majorana fermion in condensed matter systems and the Kondo effect, a sample of which can be found in, e.g., 
Refs.[\onlinecite{altland_beri,altland_beri2,altland_egger,beri,beri_cooper,eriksson_nava}]. 
In principle, our  device can be experimentally realized similarly 
to setups already employed to test the existence of Majorana modes  \cite{Mourik1003,expc2,expc3}  making use of, 
e.g., heterostructures of semiconductors
coupled to s-wave superconductors \cite{alicella} or linear junctions 
between superconductors and topological insulators \cite{fukane}.
To define an effective spin index, we regard the lead index of lead electrons as an effective isospin
index. Yet, to ease the presentation, in the paper we refer to the effective isospin 
degree of freedom simply as lead electron spin. Doing so, we show that the Kondo-type coupling of $K$ to lead electrons can be either SK like, or CK like,
depending on the specific values of the tuning parameters. 

Our design allows    for changing  in a controlled  way the 
magnitude and the sign of the electronic interaction at the superconducting island.
In the language of the Anderson impurity model Hamiltonian, which provides a reliable
low-energy description of the island, this corresponds to tuning the system across a transition from the positive-$U$, to the 
negative-$U$ regime. For both signs of $U$ the island ground state keeps twofold degenerate, 
thus triggering the onset of Kondo physics when connected to the leads. However,  the nature of the 
degenerate ground state doublet strongly depends on the sign of $U$. Eventually, this   
implies a transition from the SK to the CK regime at the change in the sign of the 
electronic interaction strength \cite{taba}.  

Within our  device, 
the SK and the CK phase do not overlap in parameter space with each other. This avoids the simultaneous presence of both effects which, though 
making the physical scenario richer, does not possibly allow for a clear-cut  
detection of the latter effect against the former one \cite{zitko}. In fact, the separate 
detection of either effect is even more favored by the fact that the SK and  the CK phase are
separated, in parameter space,  by an intermediate, ``disconnected lead'' phase, in which the impurity plays
no relevant role for the low-$T$ physics of the system. 

To probe the various phases of our system, we propose to look at the ac  conductance {\it across} the impurity,
both over the same lead (``intralead''), as well as between different leads (``interlead''). 
Eventually, we show how a synoptic  comparison of the intralead and of the interlead ac conductances 
offers a simple, though effective, way of detecting the SK and   CK effects in our system.
Moreover, when tuned within the CK phase, the Kondo impurity triggers off-diagonal conduction 
via a peculiar realization of crossed Andreev reflection between the two leads.  In analogy to 
the devices discussed in Refs.[\onlinecite{been_0,gpl}], this suggests that, in a ``dual'' setup, in which a Cooper
pair is injected into the leads through the impurity, our system might effectively work as a 
``long-distance electronic entangler,''  with potential applications to realizing large-distance entangled 
two-particle states.   

The paper is organized as follows:

\begin{itemize}

\item In Sec.\ref{modham}, we introduce the lattice model Hamiltonian for our system. In particular, we 
discuss in detail how to engineer the double superconducting island at the center of the system, so to
make it work as a  spin-Kondo, or charge-Kondo impurity,  by acting  onto a pertinent tuning parameter.

\item In Sec.\ref{effective} we derive the effective low-energy, long-wavelength, continuum Hamiltonian description of 
our system in its various phases: the spin-Kondo phase, the charge-Kondo phase, and the decoupled lead phase. 
 
 \item In Sec.\ref{rgbou}, we  resort to a perturbative renormalization group analysis, to recover how the 
 system scales with a running scale $D$ [which we identify with either the (Boltzmann constant $k$ times the) temperature $T$, or with 
 the frequency of the ac applied voltage $\omega$, depending on which scale is larger] toward the fixed point corresponding to each one of 
 its phases.
 
 \item In Sec.\ref{conductance}, we discuss  the dependence on $\omega$ 
of the intrawire and of the interwire ac conductance in each phase, focusing onto the low-temperature 
regime $\omega , k T_K \gg T$ and paying particular attention to the onset of the nonperturbative 
Kondo regime in the SK and in the CK phase. Eventually, we highlight how an appropriate 
measurement of the ac  conductances as a function of $\omega$ and of the system parameters provides an effective mean to 
map out the phase diagram of our system. 

\item In Sec.\ref{conclusions}, we summarize our result by also discussing about 
a possible practical realization of our system and by eventually highlighting possible further 
developments of our work.

\item In the various appendices, we report the mathematical details of our derivation.

\end{itemize}

\noindent

\section{Model Hamiltonian}
\label{modham}

Our device is sketched in Fig.\ref{ladder}. To model it, we resort to a lattice Hamiltonian for 
two leads,   which we represent  as two $(2 \ell + 1)$-site chains, with $\ell$ eventually sent to $\infty$. The 
 lattice Hamiltonian for the leads, $H_{\rm 0,Lat}$, is given by  
 
\beq
H_{\rm 0, Lat} =  \sum_{a = 1,2}  \{ -  J_a \sum_{ j = -\ell}^{\ell - 1 } [ c_{j, a}^\dagger c_{j+1 , a } + c_{ j + 1 , a}^\dagger c_{j , a } ]
 - \mu_a \sum_{j = -\ell }^\ell c_{j , a }^\dagger c_{j , a }  \} 
\:\:\:\: ,
\label{mha.1}
\eneq
\noindent
with  $\{ c_{j , a } , c_{j , a }^\dagger \}$ being single-fermion annihilation/creation operators at 
site-$j$ of lead $a$, obeying  the standard anticommutation relations $\{ c_{ j , a} , c_{j' , a'}^\dagger \}
= \delta_{ j , j'} \delta_{a , a'}$. $J_a , \mu_a $ respectively correspond to the single-fermion 
hopping strength and  to the on-site chemical potential in lead-$a$. For the sake of simplicity, in the following 
we choose  the parameters entering Eq.(\ref{mha.1})   independent of  $a$. 
In fact, this only quantitatively affects the final result and, in any case, one may readily discuss 
the case of different parameters in the two
leads by using the approach discussed in, e.g., Ref.[\onlinecite{giuaf_1}] in the general case of a 
ladder of interacting quantum wires.

\begin{figure}
\center
\includegraphics*[width=0.6 \linewidth]{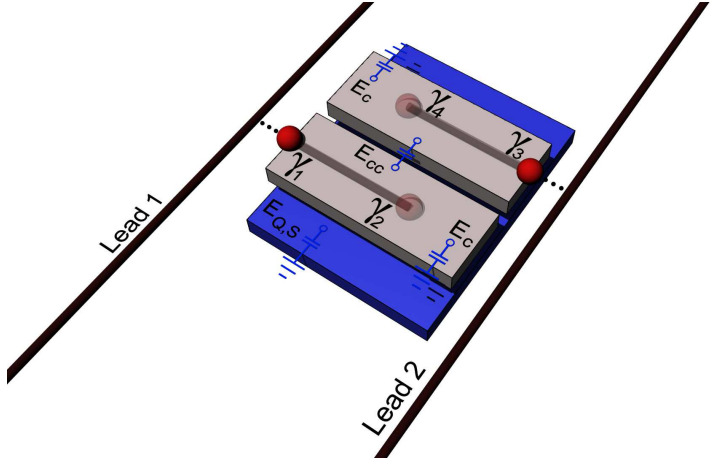}
\caption{Our proposed system: a couple of interacting spinless quantum wires is 
deposited on a  double topological island, hosting four Majorana fermions. 
The Majorana fermions $\gamma_1$ and $\gamma_3$ are tunnel-coupled to lead-1 and -2,
respectively, as highlighted by the dashed lines in the figure.} 
\label{topo_island}
\end{figure}

We now show how it is possible to realize a tunable Kondo Hamiltonian by connecting the 
two-leg ladder to a double superconducting   island (DSI), with pertinently chosen parameters.  
Specifically, we consider  an adapted version of the topological  
Kondo Hamiltonian introduced   in Ref.[\onlinecite{fu}], and then widely studied
in Refs.[\onlinecite{beri_cooper, beri, zazunov_altland,
galpin, altland_beri, altland_beri2, zazunov_levi, eriksson_mora,
eriksson_nava}]. In Fig.\ref{topo_island} we draw a sketch of the 
proposed device. It  consists of two mesoscopic s-wave superconducting islands
with two spinless quantum wires  deposited  onto each of them. According to Refs.[\onlinecite{oreg,dassarma}],
 we expect four localized real Majorana 
modes to emerge at the  end points of the wires. Let $\gamma_1 , \gamma_2$ and 
$\gamma_3 , \gamma_4$ be  the two Majorana modes arising at wire 1 and 2, 
respectively. To construct the tunable Kondo Hamiltonian, we  
assume that  $\gamma_1$ and $\gamma_3$ are tunnel-coupled to, 
respectively, lead-1 and lead-2.  
Also, we assume that  the length of each 
wire deposited on the islands and the distance  between the wires  are  large
enough to  suppress the direct tunneling between the
two Majorana modes at each wire. Yet, the two Majorana modes at the same island 
are assumed to be coupled to each other via the  capacitive  charging 
energy between  each island and the ground. Finally, we assume a nonzero 
direct  cross-capacitance coupling between the two superconducting islands and 
a Josephson coupling allowing for Cooper pair exchange between the islands and an underneath
 superconducting island $S$. As a result, the total Hamiltonian for the double-island 
system is given by  

\begin{equation}
H_{\rm Island}=H_{I,1}+H_{I,2}+H_{C}+H_{S} 
\;\;\;\; . 
\label{eq:full hamiltonian topological}
\end{equation}
\noindent
In Eq.(\ref{eq:full hamiltonian topological}), $H_{I,1}$ and $H_{I,2}$ describe 
the two islands coupled to  $S$.  They are defined so that  

\beq
H_{I,1}+H_{I, 2}  =  -E_{J}\cos ( \chi_{1} ) +E_{C}\left[2N_{1}+n_{1}-Q_{1}\right]^{2}
  -E_{J}\cos ( \chi_{2} )+E_{C}\left[2N_{2}+n_{2}-Q_{2}\right]^{2}
 \:\:\:\: , 
 \label{chaen}
\eneq
\noindent
with $n_{i}=\frac{1}{2}\left[1+i\gamma_{2i-1}\gamma_{2i}\right]$, so that 
$2N_i + n_i$ is  the 
total charge (in units of $e$) lying at island-$i$, including a possible quasiparticle 
occupying the Dirac level made out of the two Majorana modes, and   
with  $Q_{i}$ being the back-gate voltage, determined by the voltage
across the capacitor.    $E_{C}$ is the    
 charging energy of each island, and $- E_J \cos ( \chi_i)$
 corresponds to the  Josephson coupling between island-$i$ and the 
superconductor underneath, with  the   phase difference $\chi_{i}$  canonically conjugate to
the number of Cooper pairs $( N_i)$.  
\noindent
$H_C$ in Eq.(\ref{eq:full hamiltonian topological})
describes the cross-capacitive coupling between the two islands. It is given by   

\begin{equation}
H_{C}=E_{CC}\left[2N_{1}+n_{1}-Q_{1}\right]\left[2N_{2}+n_{2}-Q_{2}\right]
\;\;\;\; , 
\label{ccoupling}
\end{equation}
\noindent
with $E_{CC}$ being the cross-charging energy.  Finally, $H_S$ 
describes the superconducting island $S$, which we 
assume to be large enough for it to be able to absorb/emit Cooper 
pair at no additional cost of energy. The charging energies of each island and the
cross-charging energy are obtained from the inverse of the  Maxwell capacitance matrix, $\mathbf{C}$,
that for the system described above reads as 

\begin{equation}
\mathbf{C}=\left(\begin{array}{cc}
C_{1}+C_{12} & -C_{12}\\
-C_{12} & C_{2}+C_{12}
\end{array}\right)
\;\;\;\; .
\end{equation}
\noindent
The diagonal entries correspond to the partial capacitances relative to ground of each island, i.e., the sum of the cross 
and auto capacitances, while the off-diagonal terms are minus the 
cross capacitance between the two islands. The diagonal elements are positive, while the off-diagonal ones are negative \cite{maxwell1,maxwell2}.
From the inverse of the capacitance matrix, assuming $C_1=C_2=C$, we derive

\begin{equation}
E_{C}=\frac{e^{2}}{2}\frac{C+C_{12}}{C^{2}+2 C C_{12}}
\;\;\;\; ,
\end{equation}
\noindent
for the islands capacitance and
\begin{equation}
E_{CC}=\frac{e^{2}}{2}\frac{C_{12}}{C^{2}+2 C C_{12}}
\;\;\;\; ,
\end{equation}
\noindent
for the cross-capacitance coupling. It follows that $E_{C}>E_{CC}>0$. In the following, we further
assume that the parameters of the two mesoscopic islands 
have been chosen so that they lie within the ``charging'' regime, 
in which $E_C / E_J \gg 1$. In this case, Coulomb blockade 
prevents Cooper pairs from tunneling across the island, except 
if the backgate potential is tuned at the degeneracy point between
states with different total charge at the island. In our specific case, 
the single-fermion state associated to the pair of real Majorana modes 
can be combined into Dirac complex fermion operators,
$a_1 = \frac{1}{2} ( \gamma_1 + i \gamma_2)$, and 
$a_2 = \frac{1}{2} ( \gamma_3 + i \gamma_4 )$, thus 
allowing for low-energy charge tunneling processes across the impurity 
involving a single quasiparticle, rather than a Cooper pair. Therefore, 
setting $Q_i$ at  each island so that $Q_{i}=2\bar{N}_{i}+\frac{1}{2}$,    
with integer $\bar{N}_i$, 
allows for defining the low-energy subspace at the double junction
as spanned by the four states with $\bar{N}_i$ Cooper pairs at 
island $i$, with the mode corresponding to $a_i$ full, or empty. Listing those states, 
together with the corresponding energy eigenvalues, we obtain 
the set 
 
\begin{eqnarray}
\left|\bar{N}_{1},\bar{N}_{2},0,0\right\rangle ,\  &  & \epsilon_{0,0}=\frac{1}{2}E_{C}+\frac{1}{4}E_{CC} \; , \nonumber \\
\left|\bar{N}_{1},\bar{N}_{2},1,1\right\rangle ,\  &  & \epsilon_{1,1}=\frac{1}{2}E_{C}+\frac{1}{4}E_{CC}\; , \nonumber \\
\left|\bar{N}_{1},\bar{N}_{2},1,0\right\rangle ,\  &  & \epsilon_{1,0}=\frac{1}{2}E_{C}-\frac{1}{4}E_{CC}\; , \nonumber \\
\left|\bar{N}_{1},\bar{N}_{2},0,1\right\rangle ,\  &  & \epsilon_{0,1}=\frac{1}{2}E_{C}-\frac{1}{4}E_{CC}
\;,
\label{eq:states}
\end{eqnarray}
\noindent
with $   \left|\bar{N}_{1},\bar{N}_{2},\nu_1 ,\nu_2\right\rangle$ denoting the state
with $\bar{N}_i$ Cooper pairs at island $i$ and $\nu_i$ additional quasiparticle 
in the level determined by the Majorana modes (clearly, $\nu_i = 0 , 1$),
and with  the energies   measured with respect to a common reference level. As we
discuss in the following, the level structure summarized in Eqs.(\ref{eq:states})
is enough to induce an effective  Kondo Hamiltonian, except that, in order to 
recover both SK and   CK effect, one has to have a window of values of 
parameters corresponding to an attractive interisland interaction (that is, $E_{CC}$ has 
to become $<0$). This is the main motivation for introducing $S$,   which is  
coupled to the two island by a small Josephson  term
that allows to form a Cooper pair in the superconductor through the
annihilation of the two Dirac fermions in the islands, and vice versa.
At low energies, only particles populating the $a_i$-levels 
are involved. Therefore, the corresponding processes are   described by the hopping Hamiltonian

\beq
H_{S}  =  -\tau\left(a_{1}a_{2}e^{2i\varphi}+a_{2}^{\dagger}a_{1}^{\dagger}e^{-2i\varphi}
\right)
\:\:\:\: . 
\label{shop}
\eneq
\noindent
Differently from island-1 and -2, $S$ hosts no Majorana modes. Therefore, at low 
energies, charges can enter and exit it only as Cooper pairs. To fix the number of 
Cooper pairs at $S$, we assume that it has a finite capacitive energy 
$E_S$, which enters the corresponding Hamiltonian given by 

\beq
H_{S}^{(0)}  =  -E_{J , S} \cos\chi_{S}+E_{Q , S}\left[2N_{S}-Q_{S}\right]^{2}
\;\;\;\; . 
\label{shop.2}
\eneq
\noindent
As there is no cross-capacitance terms between the two islands and $S$, the capacitive  energy of $S$
is simply given by the inverse of its capacitance, that is $E_S=e^2 /(2C_S)$.
Tuning $Q_S = 2 \bar{N}_S$ and 
assuming $E_{Q , S} / E_{ J , S } > 1$, Coulomb blockade pins at $N_S$ 
the number of Cooper pairs at $S$. In this case, charges can tunnel from 
the islands to $S$, and vice  versa, only in virtual processes. This implies 
a corresponding lowering of the energy of the states with $\nu_1 = \nu_2 = 0$ and 
$\nu_1 = \nu_2 = 1$ by an amount that appears at second order in
$\tau$ and  is given by $\epsilon_{\tau}=-\frac{C_{S}\tau^{2}}{4e^{2}}$.
Taking this result and the level diagram in Eqs.(\ref{eq:states})
altogether, the double-island dynamics  
is described by the effective Hamiltonian

\begin{equation}
H_{\rm Island}^{\rm Eff}=
\delta\left(\left|0,0\right\rangle \left\langle 0,0\right|+
\left|1,1\right\rangle \left\langle 1,1\right|-\left|1,0\right\rangle \left\langle 1,0\right|-
\left|0,1\right\rangle \left\langle 0,1\right|\right)
\:\:\:\: .
\label{efiland}
\end{equation}
\noindent
To simplify the notation, in Eq.(\ref{efiland}) we have omitted 
the labels associated to the number of Cooper pairs on the two 
islands. The right-hand side of Eq.(\ref{efiland}) depends only 
on the tuning parameter $\delta=\frac{E_{CC}}{4}-\epsilon_{\tau}$,
which we will use as a control parameter to switch from SK to CK effect.
For the following discussion, it is useful to rewrite $H_{\rm Island}^{\rm Eff}$
in terms of the Dirac complex fermion operators $a_1 , a_2$ and of their
 Hermitian  conjugates as

\beq
H_{\rm Island}^{\rm Eff} = \delta \{ 1 - 2 [ a_1^\dagger a_1 + a_2^\dagger a_2 ] + 4 a_1^\dagger a_1 a_2^\dagger a_2 \} 
\:\:\:\: . 
\label{bint.1}
\eneq
\noindent
Formally, the coupling between the DSI  and the leads  
is  described by the tunneling Hamiltonian $H_t$, which we model 
in analogy to what is done in  Ref.[\onlinecite{fu}], as  

\begin{equation}
H_{t}=-t\sum_{a=1,2} \left(c_{0,a}^{\dagger}a_{a}+
c_{0,a}^{\dagger}a_{a}^{\dagger}e^{-i\omega_{a}}\right)+{\rm h.c.}
\:\:\:\: . 
\label{tunh.1}
\end{equation}
\noindent
In Eq.(\ref{tunh.1}), the term $c_{0,a}^{\dagger}a_{j}$ describes the transfer of a fermion
from the $i$-th island to the central site of the corresponding lead,
with the corresponding depletion of the level $a_j$. 
The term $c_{0,a}^{\dagger}a_{a}^{\dagger}e^{-i\omega_{a}}$
represents an alternative process through which a  fermion
is created in the level $a_j$ and another one is created 
in the corresponding lead along with the annihilation of a Cooper
pair in the island by the operator $e^{-i\omega_{a}}$. Noticeably, this process
induces a transition to a state with a higher number of Cooper pairs
in the islands,  which we rule out on 
projecting onto the low-energy ground state manifold of the islands. 
Therefore, we drop it henceforth from the tunneling Hamiltonian and 
describe the DSI coupled to the ladder by means of the 
boundary Hamiltonian $\hat{H}_B$ given by 

\beq
\hat{H}_B = H_{\rm Island}^{\rm Eff} -t\sum_{a =1,2} \{ c_{0, a }^{\dagger}a_{a }+
a_{a}^{\dagger}c_{0,a} \}
\:\:\:\: . 
\label{hbun}
\eneq
\noindent
In addition to the direct tunneling between the leads and the DSI, a local density-density 
interaction Hamiltonian may arise, as well. The corresponding 
 Hamiltonian  $H_{\rm DI}$ can be simply modeled as  

\beq
H_{\rm DI} = \sum_{a  , b = 1,2} \mu_{a , b } c_{0 , a}^\dagger c_{0 , a} a_b^\dagger a_b 
\:\:\:\: . 
\label{hbun.2}
\eneq
\noindent
In the following, we use the boundary Hamiltonian $H_B = \hat{H}_B + H_{\rm DI}$ to 
discuss   the  crossover between the SK and the CK regimes at the impurity by also 
pointing out the remarkable emergence of an intermediate ``disconnected lead'' (DL)
phase, with peculiar properties. 

\section{Effective impurity Hamiltonian in the various  regimes}
\label{effective}

To describe the impurity dynamics in our system, we resort to 
pertinent approximations for $\hat{H}_B$ in different windows of values of 
the various parameters. In fact, we see that  $\delta$ is the 
only scale related to  the isolated impurity. The other relevant scales 
are the tunneling strength  $t$ and the local density-density interaction 
strengths,  the $\mu_{a,b}$'s in Eq.(\ref{hbun.2}) which, consistently with 
our symmetry assumption, we choose so that   $\mu_{1,1} = \mu_{2,2} = \mu_{\rm d}$, and 
$\mu_{1,2} = \mu_{2 , 1} = \mu_{\rm od}$. A first important limit  
 corresponds to $ | \delta |   \to  \infty$.  In this limit, 
 the low-energy manifold of the system is twofold degenerate. In particular, 
  for  $\delta  \to + \infty$,  the two 
degenerate ground states correspond to the ``mini-domain walls'' of Ref.[\onlinecite{gart}], that is, 
to the $ | 1 , 0 \rangle$ and to the $ | 0 , 1 \rangle$ eigenstates of the DSI, while, for 
$\delta \to - \infty$,  the two 
degenerate states correspond  to the $ | 0 , 0 \rangle$ and to the $ | 1 , 1 \rangle$ eigenstates of the DSI.

Leaving aside, for the time being, the density-density interaction encoded in 
$H_{\rm DI}$ in Eq.(\ref{hbun.2}), we see that, 
at finite values of the $t_a$'s, tunneling processes between the degenerate ground states are 
accounted for by resorting to an effective, Kondo-type description of the interaction of the DSI with the leads. 
To do so, we employ the Schrieffer-Wolff (SW) procedure, which we  illustrate in details in 
Appendix \ref{sw}. In the symmetric case $t_1 = t_2 = t$, the leading boundary operator describing the 
residual dynamics within the low-energy subspace of the states of the DSI is either a 
SK Hamiltonian $H_{\rm K , S}$, in the case $\delta > 0$, or a CK
Hamiltonian, $H_{\rm H , C}$, for $\delta < 0$.   In particular, on applying the SW  transformation to 
$\hat{H}_B$,  one obtains 

\begin{itemize}

\item {For $\delta > 0$, the spin-Kondo (SK) Hamiltonian $H_{\rm K,S}$, given by }

\beq
 H_{\rm K , S}  = J_S \: \vec{S}_0 \cdot \vec{\cal S} 
 \:\:\:\: , 
 \label{iskham}
 \eneq
 \noindent
 with  $J_S = 2 t^2 / \delta$ and the impurity spin operator $\vec{\cal S}$ and the lead  spin density operator $\vec{S}_j$ 
 defined  as 
 
 \beq
 {\cal S}^\alpha = \frac{1}{2} \: \sum_{u,u' = 1,2} a_u^\dagger \tau^\alpha_{u,u'} a_{u'} \;\;\; , \;\;
 S_j^\alpha =  \frac{1}{2} \: \sum_{u,u' = 1,2} c_{j,u}^\dagger \tau^\alpha_{u,u'} c_{j,u'} 
 \;\;\;\; ,
 \label{ish.a}
 \eneq
 \noindent
with  $\tau^\alpha$, $\alpha = x,y,z$ being the Pauli matrices.

\item {For $\delta < 0$, the charge-Kondo (CK) Hamiltonian $H_{\rm K,C}$, given by}

\beq
 H_{\rm K , C}  = J_C \: \vec{T}_0 \cdot \vec{\cal T} 
 \:\:\:\: , 
 \label{iskham.x1}
 \eneq
 \noindent
 with $J_C = 2 t^2 / | \delta | $ and the impurity charge-isospin operator $\vec{\cal T}$ and the lead charge-isospin density operator $\vec{T}_j$ 
 respectively defined as  
  
\beq
 {\cal T}^\alpha = \frac{1}{2} \: \sum_{u,u' = 1,2} \tilde{a}_u^\dagger \tau^\alpha_{u,u'} \tilde{a}_{u'} \;\;\; , \;\;
 T_j^\alpha =  \frac{1}{2} \: \sum_{u,u' = 1,2} \tilde{c}_{j,u}^\dagger \tau^\alpha_{u,u'} \tilde{c}_{j,u'} 
 \;\;\;\; ,
 \label{ish.abis}
 \eneq
 \noindent
 with $\tilde{a}_1 = a_1 , \tilde{a}_2 = a_2^\dagger$, and $\tilde{c}_{j , 1} = c_{j , 1}$, $\tilde{c}_{j , 2} = c_{j , 2}^\dagger$.
  \end{itemize}
Turning on $H_{\rm DI}$ we see that, in the SK regime, it modifies the 
effective impurity Hamiltonian as 

\beq
 H_{\rm K , S} \to  \hat{H}_{\rm K , S}  =  J_S \: \vec{S}_0 \cdot \vec{\cal S} +  
 \left( \frac{\mu_{\rm d} - \mu_{\rm od}}{2} \right) \: S_0^z {\cal S}^z 
+ \left(  \frac{\mu_{\rm d} + \mu_{\rm od}}{2}  \right) \: \sum_{a = 1,2} c_{0,a}^\dagger c_{0 , a } 
\;\;\;\; , 
\label{ishkz.1}
\eneq
\noindent
that is, the effective isotropic Kondo Hamiltonian acquires a nonzero anisotropy 
along the $z$-direction as soon as $\mu_{\rm d} \neq \mu_{\rm od}$ plus 
a local scattering potential term, which does not substantially affect 
Kondo physics \cite{hewson}. At variance, in the complementary CK regime, 
on turning on $H_{\rm DI}$, we obtain 

\beq
 H_{\rm K , C} \to  \hat{H}_{\rm K , C}  =  J_C \: \vec{T}_0 \cdot \vec{\cal T}   
+ \left(  \frac{\mu_{\rm d} + \mu_{\rm od}}{2}  \right) \: \{ T_0^z {\cal T}^z + T_0^z + {\cal T}^z \} 
\;\;\;\; , 
\label{ishkz.2}
\eneq
\noindent
that is, again a nonzero anisotropy 
along the $z$-direction in the (charge) Kondo interaction terms, plus effective,
local field contributions coupled to both the impurity- and the itinerant-fermion 
effective charge-isospin operator at $j=0$. All the terms appearing at the right-hand side of 
both Eqs.(\ref{ishkz.1}) and (\ref{ishkz.2}) are ``standard'' contributions arising in 
the Kondo problem and, accordingly, their effect, at least in the simplest 
case of a perfectly screened spin-1/2 impurity, is basically well understood \cite{hewson}. 

It is also interesting to address in detail what happens at small values of $ | \delta |$.   
At  $\delta = 0$, the leads are fully decoupled from the impurity. At $\delta \neq 0$, 
to account for  the competition between the effects of a finite $ \delta$   and the hybridization between the  $a_a$ modes 
and the leads \cite{singim},  we  resort to an ``all-inclusive'' RG analysis,  
considering  the  RG equations for  all the running couplings associated to  
$\hat{H}_B$.  To do so, we define the dimensionless couplings as $\bar{\tau} = \left( \frac{D_0}{D} \right)^{1- d_f} 
\: \left( \frac{a t}{v} \right)$, $\bar{\mu}_{\rm d , od}  =  \left( \frac{a \mu_{\rm d , od} }{v} \right)$, 
$\bar{\delta} =  \left( \frac{D_0}{D} \right) 
\: \left( \frac{a  \delta }{v} \right)$, with $d_f = \frac{1}{2}$.  On employing a pertinently adapted 
version of the approach used in Refs.[\onlinecite{cardy_1,gst,grt}], we eventually get 
the full set of RG equations, given by (apart for irrelevant boundary terms, 
which can in principle be generated along the RG  procedure, and which we 
neglect in the following)

\begin{eqnarray}
 \frac{d \bar{\tau} (D) }{d \ln \left( \frac{D_0}{D} \right) } &=& ( 1 - d_f ) \bar{\tau} (D) + \bar{\mu}_{\rm d} (D) \bar{\tau} (D) \; ,  \nonumber \\
 \frac{ d \bar{\mu}_{\rm d} (D) }{d \ln \left( \frac{D_0}{D} \right)  } &=& - \bar{\delta}(D)  \bar{\mu}_{\rm od} (D) \; ,  \nonumber \\
 \frac{d \bar{\mu}_{\rm od} (D) }{d \ln \left( \frac{D_0}{D} \right)  } &=& - \bar{\delta} (D) \bar{\mu}_{\rm d} (D)  \; , \nonumber \\
 \frac{d \bar{\delta}(D)  }{d \ln \left( \frac{D_0}{D} \right)  } &=& \bar{\delta} (D) - \frac{ \bar{\mu}_{\rm d} (D) \bar{\mu}_{\rm od} (D)   }{4}
 \: . 
 \label{fullrg}
\end{eqnarray}
\noindent  
To infer from Eqs.(\ref{fullrg})   the condition for the  crossover between the DL and either 
the SK, or the CK, phase,  we simplify the right-hand side of Eqs.(\ref{fullrg}) by 
neglecting nonlinear terms in the various coupling strengths.  Accordingly, 
the only actually running couplings are now $\bar{\tau} ( D )$ and $\bar{\delta} ( D)$,  
given by 

\beq
\bar{\tau} ( D )=   \left( \frac{D_0}{D} \right)^{1 - d_f} \: \left( \frac{a t}{v} \right) 
\;\;\; ; \;\;
\bar{\delta} ( D ) =  \left( \frac{D_0}{D} \right)  \: \left( \frac{a \delta }{v} \right) 
\:\:\:\: . 
\label{rgan.1}
\eneq
\noindent 
The derivation of Eqs.(\ref{fullrg}) relies on the small-coupling assumption for 
the various boundary interaction strengths. This corresponds to requiring that 
  $   \bar{\tau}  (D)  < 1$, a condition that, if one considers the running coupling 
strength in Eq.(\ref{rgan.1}), only holds up to $D \sim D_*$, with 
$D_* \sim  D_0 \: \left| \frac{a t }{v} \right|^\frac{1}{1 - d_f} $. In order for the  SW transformation 
leading to the effective Kondo Hamiltonian to apply, the condition $ | t / \delta | < 1$ must hold. 
From  Eqs.(\ref{rgan.1}), one sees that  this happens at any scale if 
$| \delta | > t $ (at the reference scale  $D=D_0$). However, due to the 
nontrivial renormalization of the running parameters, the condition can also be 
satisfied if $ | \delta | < t $. To recover this condition, we note that 
$\bar{\delta} ( D )$ takes  over  $\bar{\tau} ( D)$ at a scale   $D_{\rm Cross}$,
determined by

\beq
\bar{\delta} ( D_{\rm Cross} ) = \bar{\tau} ( D_{\rm Cross} ) 
\Rightarrow  D_{\rm Cross}  = D_0 \: \left( \frac{\delta}{t} \right)^{\frac{1}{d_f}}
\:\:\:\: . 
\label{rgan.2}
\eneq
\noindent
Therefore, in order for the Kondo regime to set in, one has to have that 
$D_{\rm Cross} > D_*$, which implies the condition 

\beq
\frac{ a | \delta |}{ v} > \left( \frac{at}{v} \right)^\frac{1}{1 - d_f} 
\:\:\:\: . 
\label{rgan.x2}
\eneq
\noindent
Once the condition in Eq.(\ref{rgan.x2}) is satisfied, the impurity dynamics is either 
described by  $\hat{H}_{\rm K , S} $ in Eq.(\ref{ishkz.1}), or by  $\hat{H}_{\rm K , C} $ in 
Eq.(\ref{ishkz.2}), depending on whether $\delta > 0 $, or $\delta < 0$.

At variance, in the DL region the boundary dynamics is described, 
as we discuss in detail in Appendix \ref{pertdelta}, by the effective local
density-density interaction Hamiltonian given by

\beq
H_{\rm DL} = \kappa c_{0,1}^\dagger c_{0 , 1} c_{0 , 2}^\dagger c_{0 , 2 } + \sum_{a = 1,2} 
\lambda_a c_{ 0 , a}^\dagger c_{0 , a } 
\:\:\:\: ,
\label{pera.11x}
\eneq
\noindent
with $\kappa$ being the effective local interlead density-density interaction strength and 
$\lambda_1 , \lambda_2$ being ``residual'' intrawire single-body potential scattering 
strengths.

In the following, we use the results we derived in this section to recover, 
after resorting to an appropriate low-energy, long-wavelength continuum limit for 
the fermionic fields in the leads, a detailed RG  analysis of the 
system  boundary dynamics in the three phases discussed above.

 \begin{figure}
 \center
\includegraphics*[width=0.4 \linewidth]{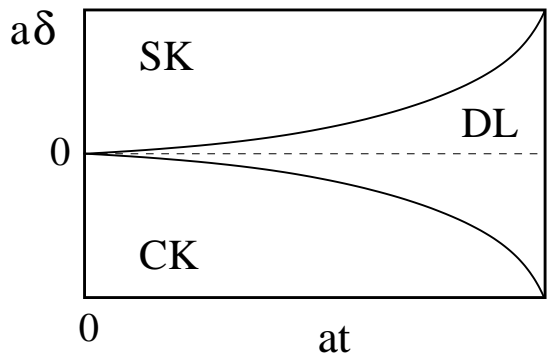}
\caption{Boundary phase diagram of our system in the $a t - a \delta $ plane: for $ | \delta | < v^{ \frac{d_f}{d_f - 1} } \: ( at )^\frac{1}{1 - d_f} $  the decoupled lead
phase sets in. 
For $\delta > v^{ \frac{d_f}{d_f - 1} } \: ( at )^\frac{1}{1 - d_f} $ and for $\delta < - v^{ \frac{d_f}{d_f - 1} } \: ( at )^\frac{1}{1 - d_f} $, 
the spin-Kondo, and the charge-Kondo  phase, respectively,  sets in. } 
\label{is_ko}
\end{figure}
\noindent

\section{Renormalization group analysis of the impurity dynamics}
\label{rgbou}

We now resort to a perturbative RG analysis, to recover the fixed point (that is, the phase) to which the 
system flows in the various regions we discuss in the previous section. To do so, we expand  the lattice fermionic fields, $c_{j , a}$, by retaining only low-energy, long-wavelength 
excitations around the Fermi points $\pm k_F = \pm {\rm arccos} \left( - \frac{\mu}{2 J} \right)$.  Therefore, 
we obtain

\beq
c_{j , a } \approx \sqrt{a} \: \{ e^{ i k_F j } \psi_{R , a} ( x_j ) + e^{- i k_F j } \psi_{L , a } ( x_j ) \}
\:\:\:\: , 
\label{exp.1}
\eneq
\noindent
with $a$ being the lattice step (which we set to 1 henceforth, except when explicitly required for 
the sake of the presentation clarity), $x_j = aj$, and $\psi_{R , a} ( x ) , \psi_{L , a} (x)$ being 
chiral fields described by the (1+1)-dimensional Hamiltonian $H_0$, given by 

\beq
H_0 = - i v \: \sum_{a = 1,2} \: \int_{-\ell}^\ell \: d x \: \{ \psi_{R , a}^\dagger ( x ) \partial_x \psi_{R , a }( x ) 
- \psi_{L , a }^\dagger ( x ) \partial_x \psi_{L , a} ( x ) \} 
\:\:\:\: . 
\label{exp.2}
\eneq
\noindent
To simplify the following derivation we note that, since we are representing the DIS as a  pointlike impurity localized 
at $x=0$, it is useful to resort to  
the ``even'' and the ``odd'' linear combinations of the chiral fermionic fields, $\psi_{e , a} ( x ) , \psi_{o , a} ( x )$, 
respectively  given by 

\begin{eqnarray}
 \psi_{e , a } ( x ) &=& \frac{1}{\sqrt{2}} \: \{ \psi_{ R ,a } ( x ) + \psi_{ L , a } ( - x ) \} \nonumber \\
 \psi_{o , a } ( x ) &=& \frac{1}{\sqrt{2}} \: \{ \psi_{R , a } ( x ) - \psi_{ L , a } ( - x ) \} 
 \:\:\:\: .
 \label{reno0.1}
\end{eqnarray}
\noindent
Apparently, $\psi_{ o , 1} ( x )$ and $\psi_{ o , 2 } ( x )$ fully decouple from 
the impurity dynamics, which is accordingly described, in the three different regions 
identified in Sec.\ref{effective}, by the boundary Hamiltonians

\begin{eqnarray}
 \tilde{H}_{\rm K , S} &=&  2 J_S \: \vec{\sigma}_e  ( 0 ) \cdot {\cal S} + 2 \left( \frac{\mu_{\rm d} - \mu_{\rm od}}{2} \right) 
 \sigma^z_e ( 0 ) {\cal S}^z + 2 \left( \frac{\mu_{\rm d} +  \mu_{\rm od}}{2} \right) \rho_e ( 0 ) \; ,  \nonumber \\
 \tilde{H}_{\rm K , C } &=& 2 J_C \: \vec{\tau}_e ( 0 ) \cdot {\cal T} +  \left( \frac{\mu_{\rm d} +  \mu_{\rm od}}{2} \right)
 \{ 2 \tau^z_e ( 0 ) {\cal T}^z + 2 \tau^z_e ( 0 ) + {\cal T}^z \} \; , \nonumber \\
 H_{\rm DL} &=& \sum_{a = 1,2} 2 \lambda_a \rho_{e,a} ( 0 ) + 4 \kappa \rho_{1 , e} ( 0 ) \rho_{2 , e} ( 0 ) 
 \; , 
 \label{reno0.2}
\end{eqnarray}
\noindent
with 

\begin{eqnarray}
 \rho_{e , a} ( 0 ) &=& \psi_{e , a}^\dagger ( 0 ) \psi_{ e ,a} ( 0 ) \; , \nonumber \\
 \rho_e ( 0 ) &=&  \sum_{a =1,2} \rho_{e , a } ( 0 ) \; , \nonumber \\
 \sigma_e^\alpha  ( 0 ) &=& \frac{1}{2} \: \sum_{u , u' = 1,2} \: \psi_{e , u}^\dagger ( 0 ) 
 \tau_{u,u'}^\alpha \psi_{ e , u' } ( 0 ) \; , \nonumber \\
 \tau_e^\alpha  ( 0 ) &=& \frac{1}{2} \: \sum_{u , u' = 1,2} \: \tilde{\psi}_{e , u}^\dagger ( 0 ) 
 \tau^\alpha_{u,u'} \tilde{\psi}_{ e , u'  } ( 0 )
 \: ,
 \label{reno0.3}
\end{eqnarray}
\noindent
and $\tilde{\psi}_{e,1} ( x ) = \psi_{e,1} ( x )$, $\tilde{\psi}_{e,2} ( x ) = \psi_{e ,2}^\dagger ( x )$. 
  [Note that the fields $\psi_{ e , a } ( x )$   contain the combinations of opposite chirality modes that, 
in each lead,  effectively couple to the DIS and can be more general than the symmetric expressions in Eqs.(\ref{reno0.1}).
Yet, for the sake of simplicity and without loss of generality in the derivation, 
in the following we employ the expressions in Eqs.(\ref{reno0.1}), which corresponds to symmetric 
coupling to the DSI of opposite chirality modes in each lead.]

In performing the RG analysis, we  
neglect the non-purely Kondo-type terms at the right-hand side of 
the first two ones of Eqs.(\ref{reno0.2}). This makes us  
  deal with a generally anisotropic Kondo Hamiltonian,  using which, in the following, we perform the RG analysis along the 
guidelines of Ref.[\onlinecite{ian_review}]. To do so, we note that, in view of the fact that only the $\psi_{e , a }$ fields do actually 
couple to the impurity spin,  the fields in the 
$o$ sector obey the continuity condition at $x= 0 $ given by 

\beq
\psi_{ o , a  } ( 0^+   ) = \psi_{  o ,  a  } ( 0^-    )  
\:\:\:\: , 
\label{reno0.6}
\eneq
\noindent
 for any value of the Kondo coupling. 
At zero Kondo coupling, the  fields in the $e$ sector 
satisfy the same boundary conditions as  in Eq.(\ref{reno0.6}), 
that is, 

\beq
\psi_{ e , a  } ( 0^+   ) = \psi_{ e ,  a  } ( 0^-   ) 
\:\:\:\: . 
\label{reno0.7}
\eneq
\noindent
  In general, we expect $\mu_{\rm d} \geq  \mu_{\rm od}$. Therefore, the RG flow of the dimensionless 
running coupling strengths in the anisotropic
(spin and charge) Kondo Hamiltonians  at the right-hand side of Eqs.(\ref{reno0.2}) is described by 
Eqs.(\ref{kt.a6}) of Appendix \ref{KT}, switching to the isotropic RG flow of Eq.(\ref{rno0}) in the 
isotropic limit. In either case, the system flows towards   the Kondo fixed point, which accordingly determines 
a change in the boundary conditions in Eq.(\ref{reno0.7}). In particular, 
at the SK fixed point one obtains \cite{ian_review}

\beq
\psi_{ e , a  } ( 0^+  ) =  e^{ - 2 i \delta_S}  \psi_{  e , a  } ( 0^-     )  
\:\:\:\: , 
\label{reno0.8}
\eneq
\noindent
with $\delta_S$ being a nonuniversal phase shift, which, at the strongly coupled fixed point, is
independent of the momentum $k$ measured with respect to the Fermi momentum of 
the chiral fermion excitations, and is equal to $\frac{\pi}{2}$ if 
particle-hole symmetry is unbroken \cite{ian_review}.  Equation (\ref{reno0.8}) simply corresponds to Nozier\`es 
Fermi-liquid boundary conditions, that is, to the fact that the formation of the 
local Kondo singlet at the impurity location prevents any other  electron 
from  accessing  that point \cite{Noz1974}. From that,  taking also into account a possible breaking of the  spin rotational 
 symmetry, one obtains that the  leading boundary operator allowed at the Kondo 
fixed point can be expressed as \cite{ian_review} 

\beq
\tilde{H}_{\rm K , S} = \alpha_S  \psi_{e , 1}^\dagger ( 0 ) \psi_{e , 1  } ( 0 )  \: 
 \psi_{e , 2  }^\dagger ( 0 ) \psi_{e , 2  } ( 0 ) 
+ \sum_{ a = 1,2} \: \beta_{S , a} \:   \psi^\dagger_{e , a  } ( 0 ) \psi_{e , a  } ( 0 )  
\:\:\:\: , 
\label{reno0.10}
\eneq
\noindent
(see also Appendix \ref{sw} for a systematic
derivation of the $\tilde{H}_{\rm K , S}$), 
with  $\psi_{ e , a} ( 0) \equiv \psi_{e , a} (0^-)$  and   $\alpha_S , \beta_{S , a }$ appropriate boundary coupling strengths.
A similar construction 
holds for the CK effect, as well,   provided one substitutes $\psi_{e , 1} ( x ) , \psi_{e , 2 } ( x )$ with 
$\tilde{\psi}_{e , 1} ( x ) , \tilde{\psi}_{e , 2} ( x )$ defined right after Eqs.(\ref{reno0.3}). As a result, 
one finds that the noninteracting 
fixed point is described by the boundary conditions at $x=0$ given by 

\begin{eqnarray}
\tilde{\psi}_{e , a } ( 0^+  ) &=& \tilde{\psi}_{e , a} ( 0^-  ) \; , \nonumber \\
\tilde{\psi}_{ o , a } ( 0^+  ) &=& \tilde{\psi}_{o , a} ( 0^- ) 
\: , 
\label{reno1.2}
\end{eqnarray}
\noindent
and that, at variance, the  CK  fixed point is described by the 
boundary conditions

\begin{eqnarray}
\tilde{\psi}_{ e , a } ( 0^+  ) &=& e^{ - 2 i \delta_C}  \tilde{\psi}_{e , a} ( 0^- ) \; ,  \nonumber \\
\tilde{\psi}_{o , a } ( 0^+ ) &=& \tilde{\psi}_{o , a} ( 0^-   ) 
\:  . 
\label{reno1.3}
\end{eqnarray}
\noindent 
Finally, one also infers that   
the leading boundary perturbation at the CK fixed point is  given by

\beq
\tilde{H}_{\rm K , C} = \alpha_C    \tilde{\psi}_{e , 1}^\dagger ( 0 )  \tilde{\psi}_{e , 1  } ( 0 )   
\:  \tilde{\psi}_{e , 2  }^\dagger ( 0 )  \tilde{\psi}_{e , 2  } ( 0 )  
+ \sum_{ a = 1,2} \: \beta_{C , a} \:  \tilde{\psi}^\dagger_{e , a  } ( 0 )  \tilde{\psi}_{e , a  } ( 0 ) 
\:\:\:\: , 
\label{reno0.13}
\eneq
\noindent
with  $\tilde{\psi}_{e , a} ( 0) \equiv \tilde{\psi}_{e , a} (0^-)$  and $\alpha_C , \beta_{C,a}$ interaction strengths. 
Remarkably, the leading boundary Hamiltonian at   both the SK and  the  CK fixed point
reported in Eqs.(\ref{reno0.10}) and  (\ref{reno0.13}), 
takes  exactly the same form as the leading boundary Hamiltonian in the DL region, the third one of 
Eqs.(\ref{reno0.2}).  As we discuss in the following, this allows for simplifying the 
derivation by making an unified analysis of the ac conduction properties of the system
at each one of the three fixed points. 

Before concluding this section, we now briefly mention the effects of 
the  local magnetic field acting on both the impurity spin and the spin of the 
conduction electrons in the effective Kondo Hamiltonians on the first two lines of 
Eqs.(\ref{reno0.2}). In fact, while, in general, a strong applied field may eventually lead 
to the suppression of Kondo effect, it is by now well   established that the Kondo effect 
instead survives the applied field, as long as  $T_K$ is 
higher than the energy scale associated to the applied magnetic field \cite{costi,grt}. 
The local magnetic field encodes the local density-density interaction at the DSI. This is 
a minor effect, which is expected to be much smaller than 
the direct electronic tunneling encoded in $\hat{H}_B$ in Eq.(\ref{hbun}). 
 This   implies that  $T_K$ is expected to be always much larger than the energy 
scale associated to the Zeeman term, which  does not 
spoil the Kondo effect.   Based on these observations, we now discuss the leading boundary 
perturbation at both the Kondo- and the DL-fixed points. As we discuss above, 
basically $H_{\rm DL}$ in Eqs.(\ref{reno0.2}) encompasses both 
the boundary Hamiltonians in Eqs.(\ref{reno0.10}) and (\ref{reno0.13}). Therefore, 
we refer to $H_{\rm DL}$ for our further analysis. As noted in Appendix \ref{KT}, a simple power counting implies 
that the dimensionless coupling associated to $\kappa$ is ${\cal K} ( D ) = \frac{D}{D_0} \kappa$. 
On lowering the running cutoff $D$, one therefore obtains that  ${\cal K} ( D )$ scales to 0 as $D / D_0$. Thus, we conclude that 
the interwire local density interaction is an irrelevant perturbation at either the SK (CK), or at the 
DL fixed point, which is consistent with the expected stability of the various 
fixed points. The left-over terms are, instead, marginal one-body scattering potential terms. These 
marginally deform the fixed point dynamics, by changing the phase shift 
 in the $e$ channel,  with minor  consequences on 
the ac  conduction properties of the system, as  we discuss in the following.

\section{ac conductance and phase diagram of the system}
\label{conductance}

In this section,   referring to the  results we derive in detail in  Appendix \ref{ac_cond}, we 
evidence how it is possible to map out the whole phase diagram of our tunable Kondo
device by means of the ac   conductance tensor and of its dependence on the frequency $\omega$.  
To perform our analysis, in the following we mostly focus onto the ac conduction properties of the 
system {\it across} the impurity,  both over the same lead, and on different leads. Specifically, 
consistently with the symmetries of our system, we discuss [in the notation of Eq.(\ref{ppl.7}) 
of Appendix \ref{pointl}], the intralead ac conductance, ${\bf G}_{(1,>);(1,<)} ( \omega )$, and 
the interlead one, ${\bf G}_{(2,>);(1,<)} ( \omega )$, as a function of $\omega$.  
As we extensively discuss in the following, whether ${\bf G}_{(2,>);(1,<)} ( \omega )$ is zero or not and, in 
this latter case, whether it takes the same sign as  ${\bf G}_{(1,>);(1,<)} ( \omega )$, or the two of 
them have opposite signs, allows us to discriminate between the various phases of our system.
  To provide a comprehensive sample of the behavior of the ac conductance tensor 
as a function of $\omega$, in the following we discuss in detail both the relevant regimes 
$\omega \gg k T_K \gg kT$ (perturbative regime) and $k T \ll \omega \ll k T_K $ (Kondo fixed-point regime).

We now discuss in detail the various regions.

\subsection{The ac conductance in the spin-Kondo phases}
\label{csc}

Within the SK-phase, the impurity dynamics is described by 
  $\hat{H}_{\rm K,S}$ in Eq.(\ref{ishkz.1}). 
 As it is not relevant for the purpose of our discussion, we 
leave aside the one-body scattering term henceforth and perform the following 
derivation by using  the anisotropic Kondo Hamiltonian $H_{\rm K,S}^{\rm anis}$,
given by 

\beq
H_{\rm K,S}^{\rm anis} = J_{S , \perp} \: \{ {\cal S}^+ \sigma^- ( 0 ) + {\cal S}^- \sigma^+ ( 0 ) \} 
+ J_{S , z} \: {\cal S}^z \sigma^z ( 0 ) 
\:\:\:\: . 
\label{csc.1}
\eneq
\noindent
Consistently with the microscopic derivation of $H_{\rm K,S}^{\rm anis}$, we expect that 
both $J_{S , \perp}$ and $J_{S , z}$ are $>0$.  Moreover, since 
$\mu_{\rm d}$ and $\mu_{\rm od}$ are both $> 0$ and one ``naturally'' expects 
$\mu_{\rm d} \geq \mu_{\rm od}$, the RG
trajectories induced by $H_{\rm K,S}^{\rm anis}$ all lie within   region I 
of the diagram in Fig.\ref{region} of Appendix \ref{KT}. According to 
Eqs.(\ref{kt.a6}) and (\ref{kt.a7}),  at a given value of the RG invariant 
 $H  = - {\cal J}_{S , z}^2  + {\cal J}_{S , \perp}^2$, with ${\cal J}_{S , \perp (z) } = \frac{a J_{S , \perp (z)}}{v}$ 
(note that, according to the above discussion, one typically gets $H<0$), 
the RG flow induced by the spin-Kondo  Hamiltonian is encoded in the running dimensionless coupling ${\cal J}_{S , \perp} ( D )$, with 
the running scale $D$ to be  identified with $\omega$ (which, as stated above,  we assume to be $\gg kT$). By integrating the second-order RG  equations for the
running coupling strength, Eqs.(\ref{kt.a6}), one obtains (trading the dependence on $D$ for an explicit dependence 
on $\omega$)

\beq
{\cal J}_{S , \perp}  ( D = \omega  ) = \frac{2 \sqrt{ | H | } \left( \frac{k T_K}{\omega} \right)^{\sqrt{| H|} }  }{1 - \left( \frac{k T_K}{\omega} \right)^{2 \sqrt{| H|}} } 
\:\:\:\: , 
\label{csc.2x}
\eneq
\noindent
with the Kondo temperature $k T_K =   D_0   \left\{ \frac{  {\cal J}_{  z} ( D_0 ) - \sqrt{| H | }}{  {\cal J}_{  z} ( D_0 ) + \sqrt{| H | }}
\right\}^\frac{1}{2 \sqrt{ |H  | }}$,    
$D_0$ being the reference energy scale (high-energy cutoff) and 
${\cal J}_{z} ( D_0 )  =   \frac{ a J_{S , \parallel} }{v}$ (note that,  in our specific case, since, in order for the 
SW transformation leading to $H_{\rm K,S}$ and to $H_{\rm K,C}$ to apply, 
we have to assume that there are no physical processes involving energies 
of the order of $| 2 \delta |$, we must properly set $D_0 = | 2 \delta |$).  

In Appendix \ref{KT} we also show how, in the isotropic limit,   $H=0$, one  obtains 

\beq
 {\cal J}_{S , \perp } ( D = \omega  ) =  {\cal J}_{S , z}  ( D = \omega )  = \frac{1}{\ln \left( \frac{k T_K}{\omega} \right)} 
\:\:\:\: , 
\label{rno0x}
\eneq
\noindent
with now $T_K$ given by 

\beq
k T_K =  D_0 e^{ - \frac{1}{{\cal J}_\perp ( D_0 ) } }
\:\:\:\: . 
\label{rno1x}
\eneq
\noindent 
So, from Eqs.(\ref{csc.2x}) and (\ref{rno0x}), we eventually conclude that, both in the anisotropic and in 
the isotropic cases, the running coupling strength is a scaling function of the dimensionless ratio 
$\omega / (k T_K)$.

In order to incorporate the nontrivial RG flow in Eqs.(\ref{csc.2x}) and (\ref{rno0x}) into 
the formulas for the ac conductances, we refer to the derivation of 
Appendix \ref{calko}. Specifically,
keeping $\omega > kT , k T_K$ and using it as  the 
running energy scale, we obtain for the running intralead and interlead ac conductances, ${\bf G}_{(1,>); (1,<)} ( \omega )$
and  ${\bf G}_{(2,>);(1,<)} ( \omega ) $,
the result in Eq.(\ref{itc.11}) of Appendix \ref{calko}, that is

\begin{eqnarray}
 {\bf G}_{(1,>);(1,<)} ( \omega  ) &=& - \frac{e^2}{2 \pi } \: \left\{ 1 - {\cal J}_{S , \perp}^2 ( \omega ) \right\} \nonumber \\
 {\bf G}_{(2,>);(1,<)} ( \omega ) &=&  - \frac{e^2}{2 \pi } 
{\cal J}_{S , \perp}^2 ( \omega ) 
 \:\:\:\: .
\label{csc.3}
\end{eqnarray}
\noindent
  As stated in Appendix \ref{calko},
the result in Eq.(\ref{csc.3}) is expected to apply from $\omega  \sim D_0$ all the way down to $\omega \sim k T_K$. 
 From the righthand side of Eq.(\ref{csc.3}) we see that turning on the Kondo coupling implies  a reduction in the  intralead
ac current   together with a nonzero interlead current.  
This is due to the peculiar features of the Kondo processes mediating the 
ac transport at nonzero 
coupling to the impurity. In terms of electron transmission across the 
impurity, transport  from, say, lead 1 to the same lead does not 
correspond to a change in the ``spin-index'' of the transmitted electron. At the onset of 
Kondo dynamics, together with the former scattering process, 
 which we depict in Fig.\ref{processes}{\bf a)} as a particle-to-particle 
transmission within lead-1, impurity spin-flip processes can induce  single electron tunneling from 
lead 1 to lead 2, mediated by the coupling to the impurity spin $\vec{\cal S}$. 
This process, which we  depict in Fig.\ref{processes}{\bf b)} as a particle-to-particle 
transmission from lead-1 to lead-2, is what is responsible for a nonzero  ${\bf G}_{(2,>);(1,<)} ( \omega )$. This is 
a  relevant process as $\omega$ goes down from  $D_0$ to $k T_K$. Accordingly, we find that 
the corresponding off-diagonal conductance takes over, when going down with $\omega$, 
until one enters the Kondo regime at the scale $k T_K$. 
An important observation is that, since the current induced in lead-2 is due to 
particle-to-particle transmission processes, it takes the same sign as 
the  current in lead-1. This is evidenced in Eq.(\ref{csc.3}) by the fact that, as long as 
the perturbative RG approach holds (that is, for $\omega  > k T_K$), 
one has that both $ {\bf G}_{(1,>);(1,<)} ( \omega  ) $ and ${\bf G}_{(2,>);(1,<)} ( \omega )$ are $<0$
(with the - sign due to our conventional definition of the positive direction for the current operators as 
the one pointing towards the impurity both from the left-  and   the right-hand sides  of the system).

On flowing towards the SK fixed point, the spin density of the lead electrons at 
$x=0$ ``locks together'' with the impurity spin, so to effectively cut the system into two 
separate parts.  Accordingly, there is 0 interlead 
ac conductance, that is, ${\bf G}_{(2,>);(1,<)} (\omega ) = 0$. Moreover, 
as the (Nozi{\`e}res-type) SK fixed point is described in terms of the single-electron 
phase shift $\delta_S$, we may employ Eq.(\ref{ppl.23}) to 
obtain  

\beq
{\bf G}_{(1,>);(1,<)} ( \omega ) = - \frac{e^2}{2 \pi} \: \cos^2 ( \delta_S ) 
\:\:\:\: .
\label{poppy.1}
\eneq
\noindent
In  the particle-hole symmetric case (corresponding to a total phase shift $\delta_S = \frac{\pi}{2}$ in the 
$e$-linear combinations of the chiral fields in each channel), Eq.(\ref{poppy.1}) implies 
${\bf G}_{(1,>);(1,<)} ( \omega )  = 0$.  
More generally, for $\delta_S \neq \frac{\pi}{2}$, one expects a reduction of 
${\bf G}_{(1,>);(1,<)} ( \omega )$  by a factor $\cos^2  ( \delta_S )$ and a simultaneous   suppression 
of ${\bf G}_{(2,>);(1,<)} ( \omega )$. 

The leading boundary perturbation at the SK fixed point is the same as the one at the 
DL fixed point, that is, $\tilde{H}_{\rm K , S}$ in Eq.(\ref{reno0.10}).  In fact, the 
term at the right-hand side of Eq.(\ref{reno0.10}) that might potentially contribute a nonzero 
${\bf G}_{(2,>);(1,<)} ( \omega )$ is the one $\propto \alpha_S$. Yet,  in analogy with  the calculations in 
the DL phase of  appendix \ref{perdis}, we find that ${\bf G}_{(2,>);(1,<)} ( \omega ) = 0$, at least to second-order 
in $\alpha_S$,  [in fact, as long as the boundary interaction describes electron scattering 
off a local singlet, one is expected to obtain ${\bf G}_{(2,>);(1,<)} ( \omega ) = 0$ at any order in $\alpha_S$]  while, to 
order $\alpha_S^2$, we obtain  

\beq
{\bf G}_{(1,>);(1,<)} ( \omega ) = - \frac{e^2}{2 \pi} \: \left\{ \cos^2 ( \delta_S) -
\frac{8 \pi \cos ( 2 \delta_S ) \alpha_S^2  \omega^2  }{3   v^2} \right\}
\:\:\:\: . 
\label{csc.5}
\eneq
\noindent
In particular, in the particle-hole symmetric case, one has 
$2 \delta_S = \pi$, which implies

\beq
{\bf G}_{(1,>);(1,<)} ( \omega ) = - \frac{e^2}{2 \pi} \: 
\frac{8 \pi  \alpha_S^2  \omega^2  }{3   v^2}   
\:\:\:\: . 
\label{csc.6}
\eneq
\noindent
To summarize, we have shown that, on lowering $\omega$ from $\omega \sim D_0$ to $\omega \sim k T_K$, 
${\bf G}_{(1,>);(1,<)} ( \omega )$ is  suppressed 
by  the  Kondo interaction. At the same time, the relevance of 
the ``effective'' spin-flip processes induces a nonzero ${\bf G}_{(2,>);(1,<)} ( \omega )$, 
which increases as $\omega$ is lowered towards $k T_K$. Since a spin-flip process here corresponds to 
a particle/hole tunneling from one lead  as an injected particle/hole to the 
other one, the ac currents induced in the two leads by means of a voltage bias applied to 
either one of them flow towards the same directions. 
Once the system has flown to the SK fixed point, it may, or may not, exhibit a finite
${\bf G}_{(1,>);(1,<)} ( \omega )$ , depending on whether particle-hole symmetry is broken, or not  \cite{afl_2}. 
The  leading   correction to the ac conductance tensor is diagonal, as well, 
and $\propto \omega^2$, consistently with Nozi{\`e}res Fermi liquid theory \cite{Noz1974}.  
 
\subsection{The ac conductance in the charge-Kondo phases}
\label{ccc}

\begin{figure}
 \center
\includegraphics*[width=0.9 \linewidth]{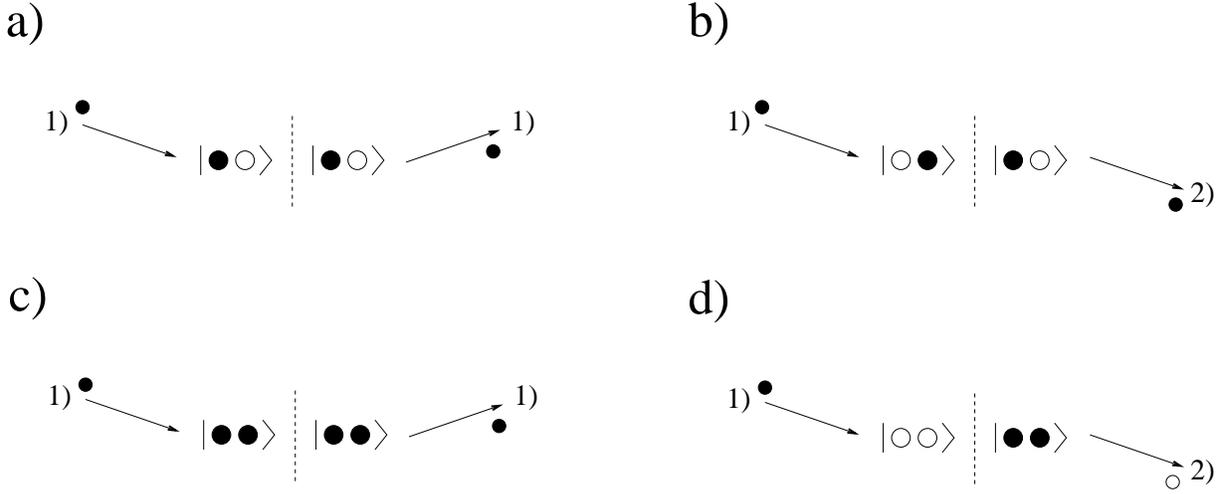}
\caption{ Sketch of the possible single-particle transmission processes that 
can take place in either the SK, or the CK , phase , if an incoming 
particle from lead-1 hits the effective magnetic impurity. The ket represents the 
``impurity'' state, so that a filled (empty) dot corresponds to the full (empty) state 
corresponding to the fermionic mode $a_1$ (left-hand dot) or $a_2$ (right-hand dot) 
in Eq.(\ref{bint.1}). In particular : \\
{\bf a)} In the SK phase, 
the particle from lead-1 is transmitted as a particle towards the same lead. The 
impurity state is $ | \Uparrow \rangle$ before, and after, the scattering process; \\
{\bf b)} Still in the SK phase, 
the particle from lead-1 is transmitted as a particle towards lead-2. The 
impurity state is $ | \Downarrow \rangle$ before the scattering process and 
switches to $ | \Uparrow \rangle$ after, consistently with total spin 
conservation;\\
{\bf c)} In the CK phase, 
the particle from lead-1 is transmitted as a particle towards the same lead. The 
impurity state is $ | \Uparrow \rangle$ before, and after, the scattering process; \\
{\bf b)} Still in the CK phase, 
the particle from lead-1 is transmitted as a hole towards lead-2 (crossed 
Andreev reflection). The 
impurity state is $ | \Downarrow \rangle$ before the scattering process and 
switches to $ | \Uparrow \rangle$ after, consistently with total charge 
conservation. Remarkably, total charge conservation forbids crossed Andreev reflection 
with the hole transmitted towards lead-1. } 
\label{processes}
\end{figure}

Within the CK phase, the impurity dynamics is described by the 
Hamiltonian $\hat{H}_{\rm K,C}$ in Eq.(\ref{ishkz.2}). Aside from the  CK coupling,  $\hat{H}_{\rm K,C}$ contains a  Zeeman-type 
coupling to a local magnetic field of both $\tau^z (0 )$ and ${\cal T}^z$. Out of these two terms, the former one provides an additional phase shift 
to single-electron scattering amplitudes which is different in different leads. Again, this just 
quantitatively affects the calculation  of the ac conductance, without invalidating the whole  RG  analysis of the Kondo interaction. 
The effects of the term $\propto {\cal T}^z$ are discussed in Appendix \ref{KT}. Here, we just mention 
that this term is not expected to  substantially affect the Kondo physics as long as 
the energy scale associated to the local magnetic field is much lower than    $k T_K$
 \cite{costi,otte,grt}.  Having stated this, one 
 therefore readily sees that the calculation of the ac  conductance in the 
CK case can be performed in perfect analogy as what we have done in Sec.\ref{csc} for 
the SK case. Yet, a fundamental difference in the results for the  ac conductance in the CK case   compared to the SK effect    
arises from the different nature of physical processes 
yielding a  nonzero ${\bf G}_{(2,>);(1,<)} ( \omega ) $ in the two cases. Indeed, while, in the CK case, interlead charge tunneling  is still
supported by an impurity spin flip, this process now corresponds to a switch between local states
with a net charge difference equal to $\pm 2 e$. Thus, charge is conserved in a single scattering 
process only modulo 2 and, in particular, interlead scattering processes are of Andreev type, with an
incoming particle from lead 1 emerging as an outgoing hole in lead 2. At variance, 
intralead scattering processes again correspond to particle-to-particle (hole-to-hole)
scattering events. In Fig.\ref{processes}{\bf c)} we depict intralead scattering 
processes in the charge-Kondo phase, while in Fig.\ref{processes}{\bf d)} we draw a 
sketch of a single ``crossed-Andreev-reflection''-like scattering event from 
lead-1 to lead-2, supporting interlead ac  transport. As a result, 
we now expect that  ${\bf G}_{(1,>);(1,<)} ( \omega ) $ and ${\bf G}_{(2,>);(1,<)} ( \omega ) $ 
 have opposite sign. Indeed,  for $ D_0 \geq \omega \geq k T_K$, one obtains  

\begin{eqnarray}
 {\bf G}_{(1,>);(1,<)} ( \omega ) &=& - \frac{e^2}{2 \pi } \: \left\{ 1 - {\cal J}_{C , \perp}^2 ( \omega ) \right\} \nonumber \\
 {\bf G}_{(2,>);(1,<)} ( \omega ) &=&   \frac{e^2}{2 \pi }   {\cal J}_{C , \perp}^2 ( \omega )
 \:\:\:\: , 
 \label{ccc.1}
\end{eqnarray}
\noindent
that is, Eqs.(\ref{itc.10}) of appendix \ref{calko}, with ${\cal J}_{C , \perp} ( \omega )$ being a scaling function of 
$\omega / ( k T_K)$ defined just as ${\cal J}_{S , \perp} ( \omega )$ of Eqs.(\ref{csc.2x}) and (\ref{rno0x}) by replacing the spin-Kondo 
couplings with the corresponding charge-Kondo ones.   As at the  SK fixed point, again,  when 
 flowing toward the  CK  fixed point, the charge-isospin density of the lead electrons at 
$x=0$ ``locks together'' with the impurity spin, so to effectively cut the system into two 
separate parts. Again, if $\delta_C$ is the corresponding intralead single-fermion 
phase shift, this implies  a reduction of  ${\bf G}_{(1,>);(1,<)} ( \omega )$  by a factor $\cos^2  ( \delta_C )$ and a simultaneous   suppression of  
${\bf G}_{(2,>);(1,<)} ( \omega )$.  The leading boundary operator at the CK  fixed point  is given by 
$\tilde{H}_{\rm K,C}$ in Eq.(\ref{reno0.13}). Just as in the spin-Kondo case, we 
therefore obtain, to order $\alpha_C^2$, that ${\bf G}_{(2,>);(1,<)} ( \omega )$ keeps $=0$, 
while  ${\bf G}_{(1,>);(1,<)} ( \omega )$ is corrected as  
 
\beq
{\bf G}_{(1,>);(1,<)} ( \omega ) = - \frac{e^2}{2 \pi} \: \left\{ \cos^2 ( \delta_C) -
\frac{8 \pi \cos ( 2 \delta_C ) \alpha_C^2  \omega^2  }{3   v^2} \right\}  
\:\:\:\: . 
\label{ccc.2}
\eneq
\noindent
To summarize the results of this section, we see that on lowering $\omega$, just as in 
the SK  case, ${\bf G}_{(1,>);(1,<)} ( \omega )$ is reduced 
by  the Kondo interaction, while the relevance of 
the ``effective'' spin-flip processes induces a nonzero ${\bf G}_{(2,>);(1,<)} ( \omega )$. Since, now,  a spin-flip process  corresponds to 
a particle/hole from one lead  as  injected as a hole/particle into the 
other one,  ${\bf G}_{(1,>);(1,<)} ( \omega )$ and 
 ${\bf G}_{(2,>);(1,<)} ( \omega )$ have opposite signs. Finite-$\omega$ contributions to  ${\bf G}_{(1,>);(1,<)} ( \omega )$ at the 
CK fixed point are  $\propto \omega^2$, again consistently with Nozi{\`e}res Fermi-liquid theory \cite{Noz1974}.

\noindent

\subsection{The ac conductance in the decoupled lead phase}
\label{cdc}
 
The DL  phase is  characterized by the  irrelevant boundary  interaction 
$H_{\rm DL}$ in Eq.(\ref{reno0.2}).  
In Appendix \ref{perdis}  we show how, in the decoupled lead phase, our system is expected to 
have    ${\bf G}_{(2,>);(1,<)} ( \omega ) = 0$ and 

\beq
{\bf G}_{(1,>) ; (1,<)} ( \omega ) \approx -   \frac{e^2}{2 \pi} \:  
\left\{ \cos^2 ( \delta_\kappa) - \frac{8 \cos ( 2 \delta_\kappa)  \kappa^2 \omega^2}{3 \pi^2 v^2}   \right\}
\:\:\:\: , 
\label{pps.5}
\eneq
 with $\delta_\kappa$ single-fermion phase shift at the DL fixed point.  
On top of the result in Eq.(\ref{pps.5}) it is also worth stressing that, since the boundary interaction 
describing the impurity throughout the decoupled lead phase is irrelevant, there is no ``Kondo-type'' 
expected crossover in this region, on lowering  $\omega$. Thus, we may eventually conclude 
that, lowering $\omega$  at fixed system's parameters, the Kondo-type  phases are dramatically different from 
the non-Kondo-type  one in that first of all the former ones are characterized by a strong dependence on 
$\omega$ of the ac conductance tensor, as $\omega$ is lowered towards $k T_K$, while the latter one just exhibits a mild 
dependence on $\omega$, and is almost not at all affected by the coupling of the leads to 
the superconducting island. Secondly, the fixed-point  properties are dramatically different, as 
well. Indeed, when lying within either one of the Kondo phases, ${\bf G}_{(1,>) ; (1,<)} ( \omega )$
is strongly reduced by the formation of the Kondo singlet at the DSI and is eventually forced to be $=0$ if 
particle-hole symmetry is not broken. At variance, at the DL fixed point,  ${\bf G}_{(1,>) ; (1,<)} ( \omega )$ is 
in general finite and only limited by possible one-body scattering potential terms due to the coupling to 
the DSI.

The  results of this section  allow for fully mapping out the phase diagram of the system by looking at the 
ac conductance tensor of the device as a function of both  $\omega$ and of the 
control parameter $\delta$, as we summarize in the following.

\subsection{The ac  transport properties and    phase diagram of the system}
\label{tunet}

Referring to the phase diagram of Fig.\ref{is_ko}, in the following we use 
as   tuning parameter $r =  \delta  v / (a t^2 )$. In particular, 
at $r = \pm 1$, the system undergoes two transitions between  either 
the SK, or the CK, phase (for $| r | > 1$)  and the DL phase (for $ | r | < 1$).
The three different phases can be well characterized 
by looking at   the ac conductances  as a  function 
of $\omega$, from $\omega \sim D_0$ all the way down across  $\omega \sim k T_K$ and below.

\begin{figure}
 \center
\includegraphics*[width=0.5 \linewidth]{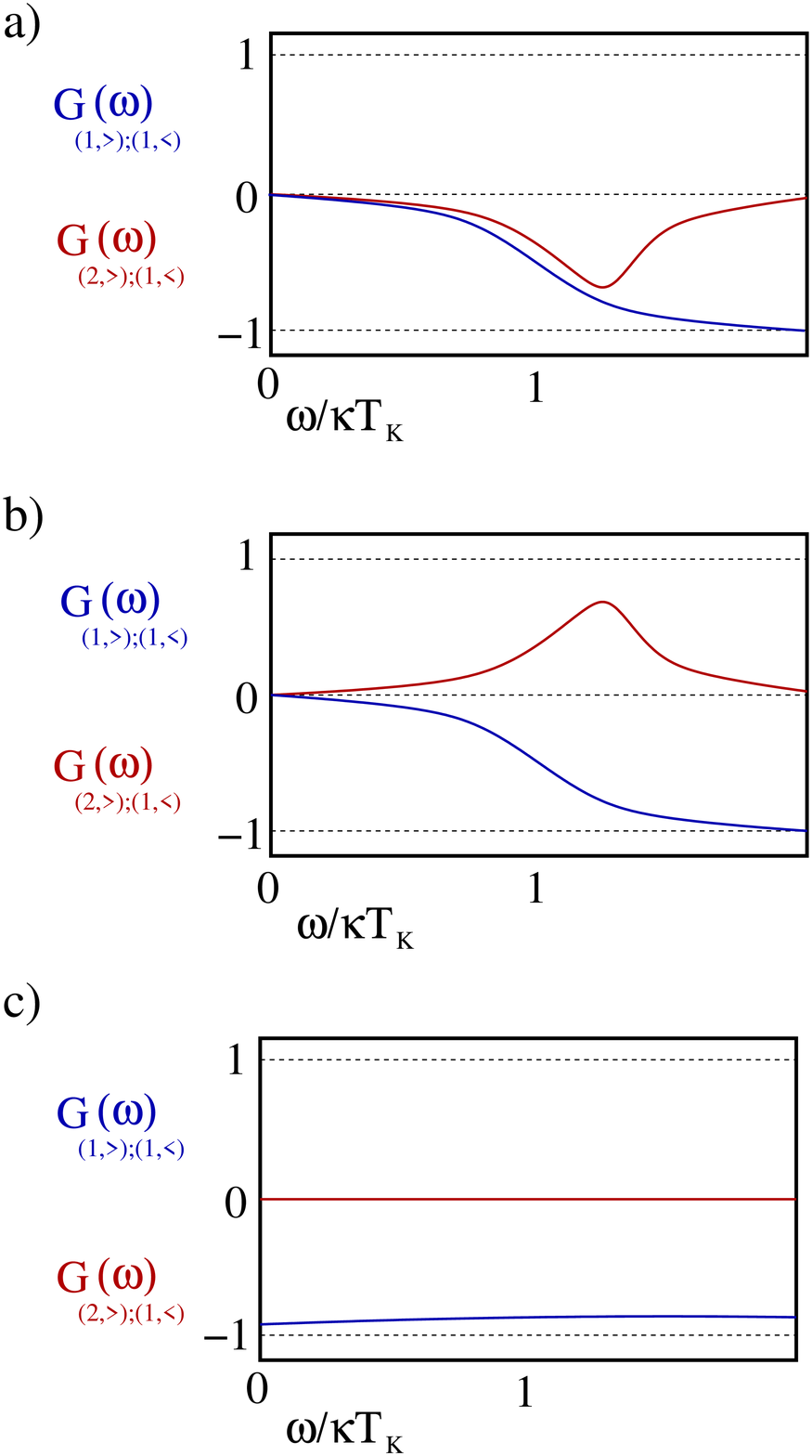}
\caption{ Sketch of   ${\bf G}_{(1,>) ; (1,<)} ( \omega )$  and of  ${\bf G}_{(2,>) ; (1,<)} ( \omega )$ as a function of $\omega$
in the various phases of the 
system, determined by different values of $r =   \delta  v / (a t^2 )$ and for $\omega , k T_K \gg kT$. From top to bottom: \\
{\bf a)} Expected behavior of ${\bf G}_{(1,>) ; (1,<)} ( \omega )$ (blue curve) and  ${\bf G}_{(2,>) ; (1,<)} ( \omega )$ (red curve) 
in units of $e^2 / (2 \pi )$ as a function of $\omega / k T_K$ in the SK region
($r>1$). The saturation of ${\bf G}_{(1,>) ; (1,<)} ( \omega )/ (e^2 / (2\pi))$ at high values of $\omega / k T_K$ may take place 
to values lower than 1, depending on the presence of potential scattering terms. 
Breaking of particle-hole symmetry may give rise to a nonzero saturation value of 
${\bf G}_{(1,>) ; (1,<)} ( \omega )/ (e^2 / (2\pi))$ as $\omega  \to 0$ (see main text for details); \\
{\bf b)} Expected behavior of ${\bf G}_{(1,>) ; (1,<)} ( \omega )$ (blue curve) and  ${\bf G}_{(2,>) ; (1,<)} ( \omega )$ (red curve) 
in units of $e^2 / (2 \pi )$ as a function of $\omega / k T_K$ in the CK  region
($r<-1$). Again, potential scattering terms can make  ${\bf G}_{(1,>) ; (1,<)} ( \omega )/ (e^2 / (2\pi))$
saturate to values lower than 1 at high values of $ \omega / k T_K$ and 
breaking of particle-hole symmetry may give rise to a nonzero saturation value of 
 ${\bf G}_{(1,>) ; (1,<)} ( \omega )/ (e^2 / (2\pi))$  as $\omega  \to 0$; \\
{\bf c)} Expected behavior of ${\bf G}_{(1,>) ; (1,<)} ( \omega )$ (blue curve) and  ${\bf G}_{(2,>) ; (1,<)} ( \omega )$ (red curve) 
in units of $e^2 / (2 \pi )$ as a function of $\omega / T_K$ in the DL region ($-1<r<1$). Due to the absence of 
a physically meaningful temperature reference scale in this region, $\omega$ has rescaled in units of 
what would be $k T_K$ for the specified values of the system's parameters in that 
region if Kondo effect were taking place.  } 
\label{sweep_T}
\end{figure}

In Fig.\ref{sweep_T} we show 
the expected behavior of ${\bf G}_{(1,>) ; (1,<)} ( \omega )$, as well as of 
 ${\bf G}_{(2,>) ; (1,<)} ( \omega )$, as a function of $\omega$  in the three phases. In the DL phase
one sees  a very mild dependence  of ${\bf G}_{(1,>) ; (1,<)} ( \omega )$
on $\omega$, with  ${\bf G}_{(2,>) ; (1,<)} ( \omega )$ being constantly = 0. At variance, 
for $ r>1$ (SK phase),  ${\bf G}_{(1,>) ; (1,<)} ( \omega )$ drops to 0 on lowering $\omega$, with a crossover scale determined by 
$k T_K$. At the same time,  ${\bf G}_{(2,>) ; (1,<)} ( \omega )$ first rises  and then drops to 0, as well, 
for $\omega < k T_K$. As we discuss above, in Sec.\ref{csc}, in this phase, ${\bf G}_{(1,>) ; (1,<)} ( \omega )$ 
 and ${\bf G}_{(2,>) ; (1,<) } ( \omega )$ have the same sign, 
in the  window of values of $\omega$ in which both are nonzero. Finally, for $r<-1$ (charge-Kondo phase)
 ${\bf G}_{(1,>) ; (1,<)} ( \omega )$ 
 and ${\bf G}_{(2,>) ; (1,<) } ( \omega )$  behave as in the spin-Kondo phase, but now, 
  in the  window of values of $\omega$ in which both are nonzero, 
they have opposite sign.

\noindent
A complementary, alternative analysis can instead be performed by 
sweeping $r$ (that is, $\delta$)  and by looking at how the interlead ac conductance 
${\bf G}_{(2,>) ; (1,<)} ( \omega )$ varies at fixed $\omega$ ($> k T_K$). As stated in  
Sec.\ref{cdc}, we expect ${\bf G}_{(2,>) ; (1,<)} ( \omega ) = 0$ within the DL phase, for $ | r | < 1$. Crossing the boundary at $ r = \pm 1 $  from within 
the DL phase, one enters   either one of the Kondo phases, in which ${\bf G}_{(2,>) ; (1,<)} ( \omega ) \neq 0$, with 
a sign depending on whether one is looking at the SK, or at the CK phase. Thus,  detecting the onset, along the 
$r$ axis, of regions with ${\bf G}_{(2,>) ; (1,<)} ( \omega ) \neq 0$,
separated by a zero interlead ac conductance, provides another mean to probe the phase diagram 
of our system. 

Finally, we note that, denoting with $T_K ( \delta )$ the Kondo temperature as a function 
of $\delta$, once all the other system parameters are fixed, since, 
according to the scaling assumption for the Kondo effect, one 
expects the conductance to be a scaling function of $ \omega / ( k T_K ( \delta ) )$ \cite{barzyk_1,barzyk_2}, increasing $ | \delta | $ (that
is, increasing $T_K ( \delta )$) is in principle equivalent to lowering $\omega$ towards $k T_K$. So, one expects plots similar to the 
ones in Fig.\ref{sweep_T} but, now, using $\delta$ as a control parameter. This  might in principle provide 
an easier way to probe Kondo scaling in our system, since   $\delta$ can actually be used as a control parameter of the 
device.  

Therefore, we may readily conclude how pertinently changing either $\omega$, or $\delta$ (or both), one can in principle
probe the whole phase diagram of the system and, in particular, the remarkable possibility 
of switching from SK, to CK effect by acting upon one control parameter only.

\section{Concluding remarks}
\label{conclusions}

In this paper, we propose how to engineer a tunable Kondo system which,  depending on 
the value of in principle one parameter only, can either work as a spin-Kondo, or as a charge-Kondo
impurity. Gauging the control parameter $\delta$, one moves the system from the  SK to the  CK phase,
passing across an intermediate, DL phase, in which the Kondo impurity is effectively 
irrelevant for the ac  conduction properties of the system.  While the main 
architecture of our device can appear rather complicated, we are confident that our 
theoretical proposal can potentially raise the interest in employing the interplay between 
emerging Majorana modes in condensed matter physics and Kondo effect to experimentally engineer 
an efficient tunable Kondo device.

Within linear response theory, we derive the intralead  and the interlead ac  conductance of the system 
as  a function of the frequency $\omega$  throughout the whole phase diagram. As a result, we show how 
the two  conductances provide an effective 
means to identify, and distinguish from each other,  the SK,   CK, and DL phases of the system.
 
To engineer our system, we employ a minimal setup, with only two spinless fermionic leads.
In principle, nowadays technology allows for realizing spinless, one-dimensional electronic conduction channels at, for instance, 
 semiconductor nanowires with a strong Rashba spin-orbit interaction and Zeeman energy \cite{sau}, as well 
 as edge states of a spin-Hall insulator \cite{bogdan}.  So, we expect it to be possible to realize our model, in 
 a realistic experiment, with spinless leads, which would rule out unwanted complications on top of the minimal 
 physics we describe here, such as onset of multichannel either SK, or CK, phases which, nevertheless, we plan to 
 study in a future work. 
 
 Finally, it is also worth recalling how, within the CK phase, our  Kondo impurity triggers interwire conduction 
via a peculiar crossed Andreev reflection between the two leads, which suggests that, in a ``dual'' setup, in which a Cooper
pair is injected into the leads through the DSI,  our system might realize an efficient 
long-distance electronic entangler.

\vspace{0.5cm}

{\bf Acknowledgements --} 

A. N. acknowledges financial support by the European Union, under ERC FIRSTORM, 
contract N. 692670.

\appendix

\section{The equilibrium ac  conductance tensor in the case of a quantum point contact between two wires}
\label{pointl}

In this appendix, we briefly review the formula for the equilibrium ac conductance tensor  in the case in 
which there is a simple structureless quantum point contact between two quantum wires, located at $x=0$, between the two wires. 
The purpose of extensively studying the simple model addressed in the following is twofold. On one hand, 
starting from the lattice model Hamiltonian and eventually resorting to the continuum formulation of 
the system allows us to define the general framework within which we compute the ac conductance tensor 
throughout our paper. On the other hand, some of the results we obtain along the derivation of this appendix 
are important to ground the discussion of the other, more complex, cases discussed in the paper. 

As a starting point, we consider the lattice model Hamiltonian given by 

\beq
H_{\rm PC ; Lat} = \sum_{a = 1,2} \: \{ - J \sum_{ j = -\ell}^{ \ell - 1} [ c_{ j , a}^\dagger c_{ j + 1 , a} + 
c_{ j + 1 , a}^\dagger c_{j , a } ] - \mu \sum_{ j = -\ell}^\ell c_{j , a}^\dagger c_{j , a } \} 
+ H_B
\:\:\:\: , 
\label{ppl.1}
\eneq
\noindent
with  the impurity  Hamiltonian $H_B$ given by 

\beq
H_B = V_d \: \sum_{ a = 1,2} c_{ 0 , a}^\dagger c_{0 , a } + V_{od} \{ c_{0,1}^\dagger c_{0 , 2 } +
c_{0,2}^\dagger c_{0,1} \} 
\:\:\:\: . 
\label{ppl.2}
\eneq
\noindent
To define the ac conductance tensor, we imagine our device to be composed of four different regions, 
$(1,<),(1,>),(2,<),(2,>)$, each one containing the sites at either the left-, or the right-hand 
side of the point contact, with the site at $j=0$ ''evenly shared'' between the two regions. 
In practice, we define the charge operator at each region so that

\begin{eqnarray}
 Q_{a , <} &=& e \sum_{ j = -\ell}^{-1 } c_{ j ,a}^\dagger c_{ j , a } + \frac{e}{2} c_{ 0 , a}^\dagger c_{0 , a } \; , \nonumber \\
 Q_{a , > } &=& e \sum_{ j = 1 }^\ell c_{ j ,a}^\dagger c_{ j , a } + \frac{e}{2} c_{ 0 , a}^\dagger c_{0 , a } 
 \:  . 
 \label{ppl.3}
\end{eqnarray}
\noindent
Defining the corresponding current operators as $I_{a , \lambda } ( t ) = \frac{d Q_{a , \lambda}  ( t ) }{d t}$, 
with $\lambda = < , >$, we consider the average current in region $a , \lambda$, ${\cal I}_{( a , \lambda)} ( t )$, 
arising when each region 
$a' , \lambda'$ is biased with a time-dependent voltage $v_{a' , \lambda' } ( t )$. Within linear 
response theory, we obtain 

\begin{eqnarray}
{\cal I}_{ ( a ,   \lambda )} ( t ) &=& \langle I_{a , \lambda} ( t ) \rangle = 
- i \: \sum_{ a' ,   \lambda'} \: \int_{-\infty}^t \: d t' \: v_{a' ,  \lambda' } ( t' ) 
\langle [ I_{a ,   \lambda} ( t ) , Q_{a' ,  \lambda'} ( t' ) ] \rangle \nonumber \\
&\equiv&  \sum_{ a' , \lambda'} \: \int_{- \infty}^\infty \: d t \: {\cal G}_{ ( a , \lambda ) ; 
( a' ,  \lambda' ) } ( t - t' ) v_{a' ,  \lambda' } ( t' ) 
\:\:\:\: , 
\label{ppl.4}
\end{eqnarray}
\noindent
with the retarded Green's function

\beq
 {\cal G}_{ ( a ,  \lambda ) ; 
(  a' , \lambda' ) } ( t - t' ) = 
- i \theta ( t - t' ) \langle [ I_{a,   \lambda} ( t ) , Q_{a' ,  \lambda'} ( t' ) ] \rangle 
\:\:\:\:. 
\label{ppl.5}
\eneq
\noindent
Resorting to Fourier space, one therefore readily obtains  the ac conductance tensor. 
To do so, one starts  from the Fourier transform of the function in Eq.(\ref{ppl.5}), given by 

\beq
{\cal G}_{  ( a ,  \lambda ) ; 
( a' ,  \lambda' ) } ( \omega ) =  \int_{- \infty}^\infty \: d t \: e^{ i \omega t} \: 
 {\cal G}_{ ( a ,   \lambda ) ; 
( a' ,  \lambda' ) } ( t  ) 
\:\:\:\: . 
\label{ppl.6}
\eneq
Denoting with ${\bf G}_{( a , \lambda) ; (a' , \lambda' )} ( \omega )$ the 
corresponding element of the ac  conductance tensor, one therefore obtains \cite{moca}

\beq
{\bf G}_{  (a ,   \lambda ) ; 
(  a' ,   \lambda' ) } ( \omega ) = 
\frac{1}{2} \: \{ {\cal G}_{  ( a ,   \lambda ) ; 
(  a' ,   \lambda' ) } ( \omega ) +  {\cal G}_{  ( a ,  \lambda ) ; 
(  a' ,  \lambda' ) } ( -  \omega )  \}
\:\:\:\: . 
\label{ppl.7}
\eneq
\noindent
Finally,  on defining the current-current retarded Green's function 

\beq
 {\cal G}^I_{ ( a ,    \lambda ) ; 
(  a' ,  \lambda' ) } ( t - t' ) = 
- i \theta ( t - t' ) \langle [ I_{a ,   \lambda} ( t ) , I_{ a' ,  \lambda'} ( t' ) ] \rangle 
\:\:\:\:, 
\label{ppl.8}
\eneq
\noindent
one readily sees, by differentiating with respect to $t$ both sides of Eq.(\ref{ppl.8}) and switching back 
to Fourier space, that one gets \cite{moca}

\begin{eqnarray}
 {\cal G}^I_{ ( a ,   \lambda ) ; 
(  a' ,   \lambda' ) } (  \omega ) &=& - i \omega  {\cal G}_{ ( a ,    \lambda ) ; 
(  a' ,  \lambda' ) } (  \omega ) - i \langle [ I_{a ,   \lambda} ( t ) , Q_{ a' , \lambda'} ( t ) ] \rangle 
\nonumber \\
&=& - i \omega  {\cal G}_{ ( a,   \lambda ) ; 
(  a' ,   \lambda' ) } (  \omega )  + {\cal G}^I_{ (a ,    \lambda ) ; 
(  a' , \lambda' ) } (  \omega = 0 ) 
\:\:\:\: , 
\label{ppl.9}
\end{eqnarray}
\noindent
with the last term at the second line  of Eq.(\ref{ppl.9}) being, in fact, independent of $t$ and 
working to cancel the $\omega = 0 $-contribution to the left-hand side.  

From the definition of the charge operators $Q_{a , \lambda}$, it is straightforward to 
derive the explicit formulas for the current operators for the lattice model Hamiltonian 
in Eq.(\ref{ppl.1}). In particular, one obtains

\begin{eqnarray}
I_{1,<} &=& - \frac{i e J}{2} \: \{ [ c_{1,1}^\dagger  - c_{-1,1}^\dagger ] c_{0,1} - c_{0,1}^\dagger 
[ c_{1,1} - c_{-1,1} ] \} -  i e V_{od}  \: \left\{ c_{0,1}^\dagger   c_{0 , 2}   - c_{0 , 2}^\dagger c_{0,1} \right\} \; , \nonumber \\ 
I_{2,<} &=& - \frac{i e J}{2} \: \{ [ c_{1,2}^\dagger  - c_{-1,2}^\dagger ] c_{0,2} - c_{0,2}^\dagger 
[ c_{1,2} - c_{-1,2} ] \} +    i e V_{od}  \: \left\{ c_{0,1}^\dagger   c_{0 , 2}   - c_{0 , 2}^\dagger c_{0,1} \right\} \; , \nonumber \\
I_{1,>} &=&   \frac{i e J}{2} \: \{ [ c_{1,1}^\dagger  - c_{-1,1}^\dagger ] c_{0,1} - c_{0,1}^\dagger 
[ c_{1,1} - c_{-1,1} ] \} -  i e V_{od}  \: \left\{ c_{0,1}^\dagger   c_{0 , 2}   - c_{0 , 2}^\dagger c_{0,1} \right\} \; ,  \nonumber \\  
I_{2,>} &=&  \frac{i e J}{2} \: \{ [ c_{1,2'}^\dagger  - c_{-1,2}^\dagger ] c_{0,2} - c_{0,2}^\dagger 
[ c_{1,2} - c_{-1,2} ] \} +  i e V_{od}  \: \left\{ c_{0,1}^\dagger   c_{0 , 2}   - c_{0 , 2}^\dagger c_{0,1} \right\} 
\:  . 
\label{ppl.10}
\end{eqnarray}
\noindent
To perform the explicit calculation of the ac conductance tensor, we resort to the expansion of    the 
lattice field operators entering Eq.(\ref{ppl.1}) in terms of chiral fermionic fields reviewed in  
Eq.(\ref{exp.1}). This  eventually leads to the continuum version of 
the system Hamiltonian, $H_{\rm PL}$, given by

\begin{eqnarray}
H_{\rm PL} &=& - i v \: \sum_{a = 1,2} \: \int_{-\ell}^\ell\: dx \: \{ \psi_{R , a}^\dagger ( x ) \partial_x \psi_{R , a }( x ) 
- \psi_{L , a}^\dagger ( x ) \partial_x \psi_{L , a} ( x ) \} + \nonumber \\
&&  V_{d } \: \sum_{a=1,2} \:   [  \psi_{R , a}^\dagger ( 0 ) + \psi_{L , a }^\dagger ( 0 )  ][ 
\psi_{R , a } (0 ) + \psi_{L , a} ( 0 )  ) ] 
+  V_{ od  } \:  \{ 
[  \psi_{R , 1}^\dagger ( 0 )  + \psi_{L , 1 }^\dagger  (0 ) ] [  \psi_{R , 2}  ( 0 )  + \psi_{L , 2 }   (0 ) ] +
{\rm h.c.} \}  
\:\:\:\: . 
\label{ppl.11}
\end{eqnarray}
\noindent
Also, in terms of the continuum fields, Eqs.(\ref{ppl.10}) become

\begin{eqnarray}
I_{1,<} &=&   e v \{ \psi_{R , 1}^\dagger ( 0 ) \psi_{R , 1} ( 0 ) - \psi_{L , 1}^\dagger ( 0 ) \psi_{L , 1 }( 0 ) \}   -  i e V_{od}  \: \left\{ 
[ \psi_{R , 1}^\dagger ( 0 ) + \psi_{L , 1}^\dagger ( 0 ) ] [ \psi_{R , 2} ( 0 ) + \psi_{ L , 2 } ( 0 ) ] -
{\rm h.c.}  \right\} \; , \nonumber \\ 
I_{2,<} &=&  e v \{ \psi_{R , 2}^\dagger ( 0 ) \psi_{R , 2} ( 0 ) - \psi_{L , 2}^\dagger ( 0 ) \psi_{L ,2 }( 0 ) \}   + i e V_{od}  \: \left\{ 
[ \psi_{R , 1}^\dagger ( 0 ) + \psi_{L , 1}^\dagger ( 0 ) ] [ \psi_{R , 2} ( 0 ) + \psi_{ L , 2 } ( 0 ) ] -
{\rm h.c.}  \right\} \; , \nonumber \\ 
I_{1,>} &=& - e v \{ \psi_{R , 1}^\dagger ( 0 ) \psi_{R , 1} ( 0 ) - \psi_{L , 1}^\dagger ( 0 ) \psi_{L , 1 }( 0 ) \}   -  i e V_{od}  \: \left\{ 
[ \psi_{R , 1}^\dagger ( 0 ) + \psi_{L , 1}^\dagger ( 0 ) ] [ \psi_{R , 2} ( 0 ) + \psi_{ L , 2 } ( 0 ) ] -
{\rm h.c.}  \right\} \; , \nonumber \\ 
I_{2,>} &=&  - e v \{ \psi_{R , 2}^\dagger ( 0 ) \psi_{R , 2} ( 0 ) - \psi_{L , 2}^\dagger ( 0 ) \psi_{L ,2 }( 0 ) \}   + i e V_{od}  \: \left\{ 
[ \psi_{R , 1}^\dagger ( 0 ) + \psi_{L , 1}^\dagger ( 0 ) ] [ \psi_{R , 2} ( 0 ) + \psi_{ L , 2 } ( 0 ) ] -
{\rm h.c.}  \right\} 
\: . 
\label{ppl.12}
\end{eqnarray}
\noindent
To further simplify the calculations, we now switch to the chiral fermionic fields $\psi_{ e , a } ( x ) , \psi_{ o , a } ( x )$
defined in Eq.(\ref{reno0.1}) of the main text. As stated before,   $\psi_{ o , 1 } ( x )$ and $\psi_{o , 2 } ( x )$ fully decouple from $H_B$ and, 
accordingly,  
they behave as free chiral fermionic fields over a segment of length $2 \ell$. The dynamics 
of $\psi_{e , 1 } ( x ) , \psi_{ e , 2 } ( x )$ is instead described by the Hamiltonian 

\beq
H_e = - i v \int_{-\ell}^\ell \: d x \: \sum_{a = 1,2} \: \psi_{e , a}^\dagger ( x ) \partial_x \psi_{e , a } ( x ) 
+ 2 V_d \sum_{a = 1,2} \psi_{e , a}^\dagger ( 0 ) \psi_{e , a } ( 0 ) + 2 V_{od} \{ \psi_{e , 1}^\dagger ( 0 ) 
\psi_{e , 2 } ( 0 ) + {\rm h.c.} \} 
\:\:\:\: . 
\label{ppl.14}
\eneq
\noindent
At the same time, the current density operators in Eqs.(\ref{ppl.12}) become

\begin{eqnarray}
I_{1,<} &=&   e v \{ \psi_{e , 1}^\dagger ( 0 ) \psi_{o , 1} ( 0 ) + \psi_{o , 1}^\dagger ( 0 ) \psi_{e , 1 }( 0 ) \}   -  2 i e V_{od}  \: \left\{ 
[ \psi_{e , 1}^\dagger ( 0 ) \psi_{e , 2} ( 0 ) -  \psi_{ e , 2 }^\dagger  ( 0 )  \psi_{e , 1} ( 0 )  \right\} \; , \nonumber \\ 
I_{2,<} &=&   e v \{ \psi_{e , 2}^\dagger ( 0 ) \psi_{o , 2} ( 0 ) + \psi_{o , 2}^\dagger ( 0 ) \psi_{e , 2 }( 0 ) \}   + 2 i e V_{od}  \: \left\{ 
[ \psi_{e , 1}^\dagger ( 0 ) \psi_{e , 2} ( 0 ) -  \psi_{ e , 2 }^\dagger  ( 0 )  \psi_{e , 1} ( 0 )  \right\} \; ,  \nonumber \\ 
I_{1,>} &=& -  e v \{ \psi_{e , 1}^\dagger ( 0 ) \psi_{o , 1} ( 0 ) + \psi_{o , 1}^\dagger ( 0 ) \psi_{e , 1 }( 0 ) \}   -  2 i e V_{od}  \: \left\{ 
[ \psi_{e , 1}^\dagger ( 0 ) \psi_{e , 2} ( 0 ) -  \psi_{ e , 2 }^\dagger  ( 0 )  \psi_{e , 1} ( 0 )  \right\} \; ,  \nonumber \\ 
I_{2,>} &=&  - e v \{ \psi_{e , 2}^\dagger ( 0 ) \psi_{o , 2} ( 0 ) + \psi_{o , 2}^\dagger ( 0 ) \psi_{e , 2 }( 0 ) \}   + 2 i e V_{od}  \: \left\{ 
[ \psi_{e , 1}^\dagger ( 0 ) \psi_{e , 2} ( 0 ) -  \psi_{ e , 2 }^\dagger  ( 0 )  \psi_{e , 1} ( 0 )  \right\} 
\:  . 
\label{ppl.15}
\end{eqnarray}
\noindent
A generic eigenmode of $H_e$ with energy eigenvalue $\epsilon$ is written in the form 

\beq
\Gamma_{\epsilon , e} = 
\sum_{a = 1,2} \: \int_{-\ell}^\ell \: d x \:   u_{e , a , \epsilon  }^* ( x ) \psi_{e , a } ( x ) 
\:\:\:\: . 
\label{ppl.16}
\eneq
\noindent
 On imposing the 
commutation relation $[\Gamma_{\epsilon , e} , H_e  ] = \epsilon \Gamma_{\epsilon , e}$, one 
obtains the Schr\"odinger equations for the corresponding wave functions  in the form 

\begin{eqnarray}
 \epsilon u_{e , 1 , \epsilon  } ( x ) & = &- i v \partial_x  u_{e ,1 , \epsilon  } ( x ) + \delta ( x ) \{ 2 V_{d }  u_{e , 1 , \epsilon } ( x ) 
 +  2 V_{od }  u_{e , 2 , \epsilon } ( x )  \} \; , \nonumber \\
 \epsilon u_{e , 2 , \epsilon  } ( x ) & = &- i v \partial_x  u_{e ,2 , \epsilon  } ( x ) + \delta ( x ) \{ 2 V_{d }  u_{e , 2 , \epsilon } ( x ) 
 +  2 V_{od }  u_{e , 1 , \epsilon } ( x )  \}  
 \: . 
 \label{ppl.17}
\end{eqnarray}
\noindent
By explicitly solving Eqs.(\ref{ppl.17}), one readily finds that the fields $\psi_{e , 1} ( 0 ) , \psi_{e , 2} ( 0 )$ 
at time $t$  can be fully expressed in terms of two sets of anticommuting energy eigenmodes $ \{ \Gamma_{\epsilon , e , \alpha } , 
\Gamma_{\epsilon , e , \beta} \} $, as 

\begin{eqnarray}
 \psi_{e , 1} ( - v t  ) &=& \frac{1}{\sqrt{2 \ell} } \: \sum_\epsilon \: e^{ i \frac{\epsilon x}{v} } 
 \: \frac{1}{2} \: \{ [ 1 + t_d ] \Gamma_{\epsilon , e , \alpha } + t_{od} \Gamma_{\epsilon , e , \beta} \} \: e^{ - i \epsilon t} \; ,  \nonumber \\
  \psi_{e , 2} ( - v t  ) &=& \frac{1}{\sqrt{2 \ell} } \: \sum_\epsilon \: e^{ i \frac{\epsilon x}{v} } 
 \: \frac{1}{2} \: \{  t_{od} \Gamma_{\epsilon , e , \alpha } + [ 1 + t_d ] \Gamma_{\epsilon , e , \beta } \} \: e^{ - i \epsilon t}
 \:  , 
 \label{ppl.18}
\end{eqnarray}
\noindent
with 

\begin{eqnarray}
 t_d &=& \frac{v^2 - V_{od}^2 + V_d^2}{( v + i V_d)^2 + V_{od}^2} \; , \nonumber \\
 t_{od} &=& \frac{- 2 i v V_{od} }{( v + i V_d)^2 + V_{od}^2} 
 \: . 
 \label{ppl.19}
\end{eqnarray}
\noindent
To compute the ac conductance tensor, we need the following retarded Green's functions

\begin{eqnarray}
 {\cal G}_a ( t - t' ) &=& - i e^2 v^2  \theta ( t - t' ) \langle [ \psi_{e , a}^\dagger ( -vt  ) \psi_{o , a } (  -vt ) + 
 \psi_{o , a}^\dagger (  -vt ) \psi_{e , a }( -vt )  , \psi_{e , a}^\dagger ( -vt'  ) \psi_{o , a} (  -vt' ) + 
 \psi_{o , a}^\dagger (  -vt' ) \psi_{e , a }( -vt' )  ] \rangle \; ,  \nonumber \\
  {\cal G}_{od} ( t - t' ) &=&     4 i e^2 V^2_{od}  \: \langle  
[ \psi_{e , 1}^\dagger ( -vt ) \psi_{e , 2} ( -vt  ) -  \psi_{ e , 2 }^\dagger  ( -vt )  \psi_{e , 1} ( -vt )  ,
 \psi_{e , 1}^\dagger ( -vt' ) \psi_{e , 2} ( -vt'  ) -  \psi_{ e , 2 }^\dagger  ( -vt' )  \psi_{e , 1} ( -vt' ) ] \rangle 
 \: . 
 \label{ppl.20}
\end{eqnarray}
\noindent
Based on the previous derivation, it is simple to explicitly compute the Fourier transform of the functions 
in Eqs.(\ref{ppl.20}). In particular, one obtains 

\begin{eqnarray}
 {\cal G}_a ( \omega ) &=&- \frac{i e^2}{4 \pi } \: \left[ 1 + {\rm Re} (t_d ) \right] \omega  = - \frac{i \omega e^2}{2\pi} \left\{ \frac{v^2 
 [ v^2 + V_d^2 + V_{od}^2] }{ ( v^2 + V_{od}^2 )^2 + 2 V_d^2 ( v^2 - V_{od}^2 ) + V_d^4} \right\}  \; ,  \nonumber \\
  {\cal G}_{od} ( \omega ) &=& - \frac{i \omega e^2}{2\pi} \left\{ \frac{2 v^2 V_{od}^2 
 }{ ( v^2 + V_{od}^2 )^2 + 2 V_d^2 ( v^2 - V_{od}^2 ) + V_d^4} \right\} 
  \: . 
  \label{ppl.21}
\end{eqnarray}
\noindent
In the $V_{od} \to 0$ limit (which is relevant for the following derivation), we obtain 
$   {\bf G}_{(2,>);(1,<)} ( \omega ) = 0$, while 

\beq
 {\bf G}_{(1,>);(1,<)} ( \omega ) =  - \frac{e^2}{2 \pi } \: \cos^2 ( \delta_S ) 
 \;\;\;\; ,
 \label{ppl.23}
 \eneq
 \noindent
 with 
 
 \beq
 \cos ( \delta_S ) = \frac{v }{\sqrt{ v^2 + V_d^2} }
 \:\:\:\: .
 \label{ppl.24}
 \eneq
\noindent

\section{Schrieffer-Wolff transformation and derivation of the effective Kondo Hamiltonian}
\label{sw}

In general, the SW procedure, when applied to a 
generic Hamiltonian $\hat{H}$, allows for recovering a reduced, effective Hamiltonian, acting on 
a limited subspace of the Hilbert space, typically determined as the subspace spanned by a certain 
set of low-lying eigenstates of $\hat{H}$. To be specific, let us consider a generic time-independent 
Schr\"odinger equation 

\begin{equation}
\hat{H}\left|\Psi\right\rangle =E\left|\Psi\right\rangle 
\:\:\:\: , 
\label{swp.1}
\end{equation}
\noindent
and suppose we want to ``project'' it onto a pertinently defined low-energy
subspace ${\bf G}$. Let ${\cal P}_{{\bf G}}$ be the projector
on ${\bf G}$. To lowest order in the ``off-diagonal'' matrix elements
connecting ${\bf G}$ to its orthogonal subspace, we obtain

\noindent 
\begin{eqnarray}
 &  & {\cal P}_{{\bf G}}\hat{H}{\cal P}_{{\bf G}}\left\{ {\cal P}_{{\bf G}}\left|\Psi\right\rangle \right\} +
 {\cal P}_{{\bf G}}\hat{H}\left[{\bf I}-{\cal P}_{{\bf G}}\right]\{\left[{\bf I}-{\cal P}_{{\bf G}}\right]\left|\Psi
 \right\rangle \}=E\left\{ {\cal P}_{{\bf G}}\left|\Psi\right\rangle \right\} \; , \nonumber \\
 &  & \left[{\bf I}-{\cal P}_{{\bf G}}\right]\hat{H}\left[{\bf I}-{\cal P}_{{\bf G}}\right]\left\{ {\cal P}_{{\bf G}}\left|
 \Psi\right\rangle \right\} +\left[{\bf I}-{\cal P}_{{\bf G}}\right]\hat{H}{\cal P}_{{\bf G}}\left\{ {\cal P}_{{\bf G}}\left|
 \Psi\right\rangle \right\} =E\left\{ \left[{\bf I}-{\cal P}_{{\bf G}}\right]\left|\Psi\right\rangle \right\} 
 \: . 
 \label{sw.2}
\end{eqnarray}
\noindent
Putting together Eqs.(\ref{sw.2}), one eventually obtains the ``projected''
Schr\"odinger equation

\noindent 
\begin{equation}
 \left\{ {\cal P}_{{\bf G}}\hat{H}{\cal P}_{{\bf G}}+{\cal P}_{{\bf G}}\hat{H}\left[{\bf I}-{\cal P}_{{\bf G}}\right]\left[E-\left[{\bf I}-
{\cal P}_{{\bf G}}\right]\hat{H}\left[{\bf I}-{\cal P}_{{\bf G}}\right]\right]^{-1}\left[{\bf I}-
{\cal P}_{{\bf G}}\right]\hat{H}{\cal P}_{{\bf G}}\right\} {\cal P}_{{\bf G}}\left|\Psi\right\rangle =
E{\cal P}_{{\bf G}}\left|\Psi\right\rangle 
\:\:\:\: . 
\label{swp.3}
\end{equation}
\noindent
 Dividing the Hamiltonian as the sum of a non perturbed contribution
plus a perturbation term $H=H_{0}+H_{1}$, we can project it onto
the ($n$ times degenerate) ground-state subspace of $H_{0}$

\noindent 
\begin{equation}
H_{0}\left|\psi_{i}\right\rangle =E_{0,i}\left|\psi_{i}\right\rangle ,\ \ \ i=1,...,n
\:\:\:\: . 
\label{swp.4}
\end{equation}
\noindent
to obtain the Brillouin-Wigner perturbation expansion

\noindent 
\begin{equation}
H_{Eff}=\sum_{i,j}h_{i,j}\left|\psi_{i}\right\rangle \left\langle \psi_{j}\right| 
\:\:\:\: ,
\label{sqp.5}
\end{equation}
\noindent
with

\noindent 
\begin{equation}
h_{i,j}=\left\langle \psi_{i}\right|H_{0}\left|\psi_{j}\right\rangle +\left\langle \psi_{i}\right|H_{1}
\left|\psi_{j}\right\rangle +\sum_{k}\frac{\left\langle \psi_{i}\right|H_{1}\left|\varphi_{k}\right\rangle \left\langle
\varphi_{k}\right|H_{1}\left|\psi_{j}\right\rangle }{E_{0}-E_{k}}
\:\:\:\: , 
\label{eq:Brillouin-Wigner}
\end{equation}
\noindent 
where the sum over $k$ runs on the low energy excited states of the
unperturbed Hamiltonian.  Equations (\ref{swp.1}) and (\ref{sw.2}) define a systematic procedure. A straightforward implementation 
of the procedure we illustrate here, allows for recovering the 
effective Kondo Hamiltonians in Eqs.(\ref{iskham}) and (\ref{iskham.x1}).

Another effective use of the  SW  transformation leads, to the residual  interaction at both the SK, and the 
CK fixed points.   To illustrate our derivation,  we consider the 
 SK fixed point. For the sake of generality, we consider an anisotropic lattice version of the lattice 
 SK model Hamiltonian in the form  
 
 \beq
 H_{\rm Lat; S} = \sum_{a = 1, 2} \{ - J \sum_{j = -\ell}^{\ell -1} [ c_{j , a}^\dagger c_{ j+ 1, a} +   c_{ j+ 1, a}^\dagger  c_{j , a} ] 
 - \mu \sum_{j = -\ell}^{\ell }    c_{j , a}^\dagger c_{ j , a}  \} +  J_{S , \perp} \: \{ S_0^+ {\cal S}^- + S_0^- {\cal S}^+ \} + 
J_{S , z} S_0^z {\cal S}_1^z 
\:\:\:\: . 
\label{kfp.a1}
\eneq
\noindent
The ground-state  at the SK 
fixed point  minimizes the boundary interaction energy encoded in 
the Hamiltonian in Eq.(\ref{kfp.a1}) as $J_{S , \perp} , J_{S , z} \to \infty$.   
To explicitly provide such a state, 
in the following,  we denote by $ | \Uparrow \rangle , 
| \Downarrow \rangle$ the two eigenstates of ${\cal S}^z$, with $ | 0 \rangle , | \uparrow \rangle , | \downarrow \rangle , 
| \downarrow \uparrow \rangle$ the states in which, respectively,  the site $j=0$ is empty in both chain, is 
filled with one electron on chain 1 and  empty on chain 2, is 
filled with one electron on chain 2 and   empty on chain 1, is filled with one electron in both chains. As a result 
of the hybridization with the impurity spin, the  locally hybridized states, which we list below together with 
the corresponding energies, are generated: 

\begin{eqnarray}
&&{\rm Local \: singlet} \: \: : \:  | S \rangle = \frac{1}{\sqrt{2}} \: \{ | \downarrow , \Uparrow \rangle - | \uparrow , \Downarrow \rangle \} 
\;\;\; ; \;\; (\epsilon_S = - \frac{1}{4} ( S_{S , z} + 2 J_{S , \perp} ) ) \nonumber \\
&& {\rm Local \: doublet } \: D \: \: : \: | D_\sigma \rangle = | 0 , \sigma \rangle \;\;\; ; \;\; (\epsilon_D = 0 ) \nonumber \\
&& {\rm Local \: doublet } \: \tilde{D} \: \: : \:  | \tilde{D}_\sigma \rangle = | \downarrow \uparrow  , \sigma \rangle \;\;\; ; \;\; ( \epsilon_{\tilde{D}} = 0 ) \nonumber \\
&& {\rm Local \: triplet } \: T_1 \: \: : \: | T_1 \rangle = |  \uparrow  , \Uparrow  \rangle \;\;\; ; \;\; (\epsilon_{T ,   1} = \frac{J_{S , z} }{4} )  \nonumber \\
 && {\rm Local \: triplet } \: T_{-1} \: \: : \:  | T_{-1} \rangle = |  \downarrow  , \Downarrow  \rangle  \;\;\; ; \;\; 
 (\epsilon_{T ,   - 1} =  
 \frac{J_{S , z} }{4} )  \nonumber \\
&& {\rm Local \: triplet } \: T_{0} \: \: : \:   | T_0  \rangle = \frac{1}{\sqrt{2}} \: \{ | \downarrow , \Uparrow \rangle + | \uparrow , \Downarrow \rangle \} 
\;\;\; ; \;\; ( \epsilon_{T , 0 } = \frac{1}{4} ( - J_{S , z} + 2 J_{S , \perp} ) )
\:\:\:\: . 
\label{kfp.8}
\end{eqnarray}
\noindent
At large values of   $J_{S , \perp} , J_{S , z}$, the system lies within the $ | S \rangle $ local state, with 
the higher-energy states in Eqs.(\ref{kfp.8}) playing a role in the allowed physical processes only as virtual states. Taking this 
into account, one can therefore go through a systematic SW  transformation,  by using as ``unperturbed'' 
Hamiltonian ${\cal H}_0 = \sum_X \epsilon_X | X \rangle \langle X |$, with $ \{ | X \rangle \}$ being the set of states listed in 
Eqs.(\ref{kfp.8}), and as ``perturbing'' Hamiltonian ${\cal H}_t$, describing the coupling between site-0 and sites-$\pm 1$  of each lead, and 
given by 

\beq
{\cal H}_t = t \{  [ c_{1,1} + c_{-1,1} ]  c_{0,1}^\dagger + [ c_{1,2} + c_{-1,2} ]  c_{0,2}^\dagger -
[  c_{1,1}^\dagger + c_{-1,1}^
\dagger ]  c_{0,1} - [  c_{1,2}^\dagger + c_{-1,2}^\dagger ]  c_{0,2} \} 
\:\:\:\: . 
\label{kfp.9}
\eneq
\noindent
As a result, one  finds that the first nontrivial boundary interaction operator, $\tilde{H}_{K,S}$,  arises to fourth-order in 
${\cal H}_t$, and is given by 

\beq
\tilde{H}_{K,S} = \sum_{ \{X , X' \}  = \{ D_\sigma , \tilde{D}_\sigma \} }
\: \sum_{ \{Y \} = \{ T_M \} } \: \left[  {\cal P}_S \frac{{\cal H}_t }{\epsilon_S - \epsilon_X} {\cal P}_X 
\frac{{\cal H}_t }{\epsilon_S - \epsilon_Y } {\cal P}_Y \frac{{\cal H}_t }{\epsilon_S - \epsilon_{X'} }
{\cal P}_{X'} {\cal H}_t  {\cal P}_S \right]
\:\:\:\: , 
\label{kfp.10}
\eneq
\noindent
 with ${\cal P}_X$ being the projector onto the space spanned by the local state $ | X \rangle$.
Plugging into Eq.(\ref{kfp.10}) the explicit expressions for ${\cal H}_t$ and for the 
various energies, one finds 

\begin{eqnarray} 
\tilde{H}_{K,S} &=&  - \zeta_{ I , \perp } \{ [  ( c_{1 , 2} + c_{-1,2} )     ,  (   c_{1,1}^\dagger  + c_{-1,1}  ) ] [ 
( c_{1,1} + c_{-1,1}  )  , (   c_{1,2}^\dagger  + c_{-1,2}^\dagger )  ]  \nonumber \\
&+&  [  (  c_{1,1} + c_{-1,1}  )  ,  ( c_{1 ,2}^\dagger  + c_{-1,2}^\dagger )  ] [(  c_{1 , 2} + c_{-1,2} )  ,
(  c_{1,1}^\dagger  + c_{-1,1}  ) ] \} \nonumber \\
 &-& \zeta_{I , z} \{ [ ( c_{1,2} + c_{-1,2} )  , ( c_{1,2}^\dagger + c_{-1,2}^\dagger )  ] - [ ( c_{1,1} + c_{-1,1} )  ,
 (  c_{1,1}^\dagger  + c_{-1,1}^\dagger ) ] \}^2 
 \;\;\;\; , 
 \label{kfp.17}
 \end{eqnarray} 
 \noindent
 with 
 
 \begin{eqnarray}
 \zeta_{I , \perp} &=& \frac{t^4}{4 \epsilon_S^2 [ \epsilon_{T_1} - \epsilon_S]} \nonumber \\
 \zeta_{I , z}  &=& \frac{t^4}{4 \epsilon_S^2 [ \epsilon_{T_0} - \epsilon_S]} 
 \;\;\;\; , 
 \label{kfp.18}
 \end{eqnarray}
\noindent
$\tilde{H}_{K,S}$ in Eq.(\ref{kfp.17}) is the lattice version of the 
operator describing the leading perturbation at Nozi{\`e}res Fermi-liquid 
fixed point as derived in, e.g., Appendix D of Ref.[\onlinecite{afl_2}]. 
Resorting to the continuum field framework by means of, 
e.g., the low-energy, long-wavelength expansions in Eqs.(\ref{exp.2}) and (\ref{reno0.1})
and inserting the continuum fields, supplemented with the 
appropriate fixed point boundary conditions, in Eq.(\ref{kfp.18}), one
recovers the continuum formula for the 
leading boundary perturbation at the SK  fixed point [Eq.(\ref{reno0.10})]. 
Similarly,  at the CK  fixed point, one 
obtains  Eq.(\ref{reno0.13}) of the main text.

\section{Derivation of $H_{\rm DL}$ in the disconnected lead phase} 
\label{pertdelta}

In this appendix, we briefly discuss the derivation of $H_{\rm DL}$ in Eq.(\ref{pera.11x}) as the 
leading boundary interaction describing the impurity within the DL phase. 
To do so, 
we start by ``artificially'' introducing two independent parameters in the DSI Hamiltonian, 
which we accordingly rewrite as 

\beq
\hat{H}_{\rm Island} = - 2 \delta_1 \: \sum_{a = 1,2} a_a^\dagger a_a + 
4 \delta_2 a_1^\dagger a_1 a_2^\dagger a_2
\:\:\:\: .
\label{pera.1}
\eneq
\noindent
Clearly, in order to recover physically meaningful results, one has to eventually set 
$\delta_1 = \delta_2 = \delta$. However, Eq.(\ref{pera.1}) comes out to be useful in that it allows 
for exactly accounting for at least part of the $\delta$-depending terms in $H_{\rm Island}$. 

Identifying $\delta_2$ as our perturbative parameter, we note that, leaving aside the density-density 
interaction (that is, setting $\mu_{\rm d} = \mu_{\rm od} = 0$), the system Hamiltonian can be written 
as  $\hat{H}_0 = \sum_{a = 1,2} \hat{H}_a$, with 

\beq
\hat{H}_a = - J \: \sum_{ j = -\ell}^{\ell  - 1} \: \{ c_{j , a}^\dagger c_{j + 1 , a } + c_{ j + 1 , a}^\dagger c_{j , a } \} 
- \mu \: \sum_{ j = - \ell}^\ell  \: c_{j , a}^\dagger c_{ j , a } - t \{ c_{0 , a}^\dagger a_a + 
a_a^\dagger c_{0 , a} \} - 2 \delta_1 a_a^\dagger a_a
\:\:\:\: . 
\label{pera.2}
\eneq
\noindent
The right-hand side of Eq.(\ref{pera.2}) contains a quadratic Hamiltonian defined over 
an $( 2 \ell + 2)$-site lattice. This can be exactly diagonalized by means of the eigenmodes $\Gamma_{ k , a }$, 
defined as 

\beq
\Gamma_{ k , a} = \sum_{ j = - \ell}^\ell \:  u_{j , a , k } c_{j , a } + \xi_{ k , a } a_a
\:\:\:\: , 
\label{pera.3}
\eneq
\noindent
and (on imposing periodic boundary conditions at $j = \pm \ell$) the wave functions satisfying the lattice 
Schr\"odinger equation

\begin{eqnarray}
\epsilon_k  u_{ j , k , a } &=& - J \{ u_{ j + 1 , k , a } + u_{ j - 1 , k , a } \} - \mu u_{ j , k , a }  \;\;\; , \;\; ( j \neq 0 ) \; , 
\nonumber \\
\epsilon_k u_{ 0 , k , a } &=& - J \{ u_{  1 , k , a } + u_{  - 1 , k , a } \} - \mu u_{0 , k , a }   - t \xi_{ k , a } \; ,  \nonumber \\
\epsilon_k \xi_{ k , a } &=& - 2 \delta_1  \xi_{k , a } - t u_{ 0 , k ,  a }  
\:  . 
\label{pera.4}
\end{eqnarray}
\noindent  
To solve Eqs.(\ref{pera.4}), we make the ansatz 

\begin{eqnarray}
u_{j , k , a } &=& \alpha_{k , a}^< e^{ i k j } + \beta_{ k , a }^< e^{ - i k j } \;\;\; , \;\; ( - \ell \leq j < 0 ) \nonumber \\
u_{j , k , a } &=& \alpha_{k , a}^> e^{ i k j } + \beta_{ k , a }^> e^{ - i k j } \:\:\: , \:\: ( 0 < j \leq \ell) 
\:\:\:\: , 
\label{pera.5}
\end{eqnarray}
\noindent
which yields $\epsilon_k = - 2 J \cos ( k ) - \mu$ and, in addition, the conditions at $j = 0$ encoded in 
 
\begin{eqnarray}
u_{ 0 , k , a } &=& \alpha_{ k , a }^< + \beta_{ k , a }^< \nonumber \\
u_{0 , k , a } &=& \alpha_{ k , a}^> + \beta_{k , a}^> \nonumber \\
 2 J \cos ( k )  \: u_{ 0 , k , a } &=& J \{ \alpha_{ k , a}^< e^{ - i k } + 
 \beta_{ k , a}^< e^{ i k }  + \alpha_{ k , a}^> e^{ i k } + 
 \beta_{ k , a }^> e^{ - i k } \} + t \xi_{ k , a } \nonumber \\
 \{ 2 J \cos ( k ) + \mu - 2 \delta_1 \} \xi_{ k , a } &=& t u_{ 0 , k , a } 
 \:\:\:\: . 
 \label{pera.6}
 \end{eqnarray}
 \noindent 
Once the system in  Eqs.(\ref{pera.6}) has been explicitly solved,   at a given $k$, 
one obtains  
 
\begin{eqnarray}
  u_{0 , k , a }   &=& \alpha \: \left\{ - \frac{ 2 i J ( \epsilon_k + 2 \delta_1 ) \sin ( k ) }{ 
  ( \epsilon_k + 2 \delta_1 ) ( \epsilon_k + \mu ) + 2 i e^{ i k } J ( \epsilon_k + 2 \delta_1 ) - t^2 } \right\} \nonumber \\
  \xi_{k , a} &=&  \alpha \: \left\{   \frac{ 2 i J t \sin ( k ) }{ 
  ( \epsilon_k + 2 \delta_1 ) ( \epsilon_k + \mu ) + 2 i e^{ i k } J ( \epsilon_k + 2 \delta_1 ) - t^2 } \right\} 
\:\:\:\: , 
\label{pera.7bis}
\end{eqnarray}
\noindent 
with $\alpha$ being an over-all normalization constant.    Considering specific solutions obeying 
scattering boundary conditions, such as 

\beq
u_{j , k , a }^{(1)} = \Biggl\{ \begin{array}{l} \alpha \{ e^{ i kj} + r_{k ,   a}^{(1)} e^{ - i k j } \} \;\;\; , \; (j<0 ) \\
\alpha t_{k , a}^{(1)} e^{ i k j } \;\;\; , \; (j>0 ) \end{array} 
\;\;\;\; , 
\label{perc.1}
\eneq
\noindent
and 

\beq
u_{j , k , a }^{(1)} = \Biggl\{ \begin{array}{l} \alpha  t_{k ,   a}^{(2)} e^{ - i k j }  \;\;\; , \; (j<0 ) \\
\alpha \{ e^{ - i kj} + r_{k ,   a}^{(2)} e^{  i k j } \} \;\;\; , \; (j>0 ) \end{array}  
\:\:\:\: , 
\label{pera.ter2}
\eneq
\noindent
 one obtains 
 
 \beq
t_{k,a}^{(1)} = t_{k , a}^{(2)}     
=   \left\{ - \frac{ 2 i J ( \epsilon_k + 2 \delta_1 ) \sin ( k ) }{ 
  ( \epsilon_k + 2 \delta_1 ) ( \epsilon_k + \mu ) + 2 i e^{ i k } J ( \epsilon_k + 2 \delta_1 ) - t^2 } \right\}
\:\:\:\: . 
\label{pera.ter3}
\eneq
\noindent
At the Fermi level one therefore obtains 

\beq
t_{k_F , a}^{(1)} = t_{k_F , a}^{(2)} = \frac{2 i \delta_1 v}{ 2 i \delta_1 v - t^2} 
\:\:\:\: , 
\label{pera.ter4}
\eneq
\noindent
with $v$ being the Fermi velocity. 
Also, given the relations
 
\beq
 c_{ 0 ,  a } = \sum_k u_{ 0 , k , a}^* \Gamma_{ k , a}  \;\;\;  ,\;\;
 a_a =  \sum_k \xi_{k , a}^* \Gamma_{ k , a}
 \:\:\:\: , 
 \label{pera.9}
\eneq
\noindent
one notes that, approximating $t_{k , a}^{(1,2)}$ with $t_{k_F , a}^{(1,2)}$ in Eq.(\ref{pera.ter4}) (which 
clearly applies to energy scales lower than $ |t|$) and 
if $\delta_1 \neq 0$, one obtains  

\beq
a_a \approx - \frac{t}{2 \delta_1} c_{0 , a} 
\Rightarrow a_a^\dagger a_a \approx \frac{t^2}{4 \delta_1^2} c_{ 0 , a }^\dagger c_{ 0 ,a } 
\:\:\:\: . 
\label{pera.10}
\eneq
\noindent 
 Equation (\ref{pera.10})  implies  $a_a^\dagger a_a \propto c_{0 , a}^\dagger c_{0 , a }$. Therefore, the whole 
impurity interaction Hamiltonian can be traded for a density-density interaction 
one, $H_{\rm DL}$, of the form 

\beq
H_{\rm DL} = \kappa c_{0,1}^\dagger c_{0 , 1} c_{0 , 2}^\dagger c_{0 , 2 } + \sum_{a = 1,2} 
\lambda_a c_{ 0 , a}^\dagger c_{0 , a } 
\:\:\:\: ,
\label{pera.11}
\eneq
\noindent
with $\kappa , \lambda_1 , \lambda_2$ parameters corresponding to 
the interwire local density-density interaction and to the residual 
intrawire local one-body potentials and $\kappa \approx  \left( \frac{ t  }{ 2  \delta_1  } \right)^4 \: \delta_2 $. 
$H_{\rm DL}$ in Eq.(\ref{pera.11}) is the Hamiltonian we used in the main text 
to discuss the effective impurity dynamics in the DL region of 
the phase diagram.

\section{Renormalization group equations for the running coupling in the effective impurity 
Hamiltonians}
\label{KT}

In this appendix, we concisely review the derivation and the solution of the   RG  equations for 
the various boundary Hamiltonians describing the impurity dynamics in the various regions of the system. 

\begin{figure}
 \center
\includegraphics*[width=0.5 \linewidth]{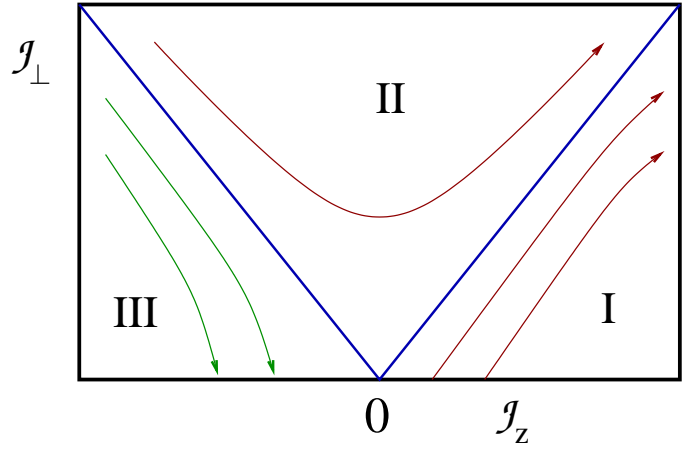}
\caption{Sketch of the RG  trajectories for the running couplings 
$ {\cal J}_\perp (D)$ and $ {\cal J}_z ( D )$.} 
\label{region}
\end{figure}

To begin with, we consider 
$ \hat{H}_{\rm K , S}$ in Eq.(\ref{ishkz.1}) and $ \hat{H}_{\rm K , C}$ in 
Eq.(\ref{ishkz.2}). For the sake of the discussion, here we leave aside purely marginal terms (that is, one-body potential 
scattering terms), whose nonuniversal effects we account for when actually computing 
the dc conductance tensor of the system. At the same time, we neglect  local fields acting on 
the effective spin impurity, whose effects we briefly discuss by the end of this appendix. 
Accordingly, in order to  encompass in our analysis both cases,  
we henceforth  consider the RG equations for the 
running couplings associated a generic  anisotropic Kondo Hamiltonian 
$H_{\rm An}$,  given by

\beq
H_{\rm An} = J_\perp \{ \psi_1^\dagger ( 0 ) \psi_2 ( 0 ) {\cal S}^- + \psi_2^\dagger ( 0 ) 
\psi_1 ( 0 ) {\cal S}^+ \} + J_z \left\{ \frac{\psi_1^\dagger ( 0 ) \psi_1 ( 0 ) - \psi_2^\dagger ( 0 ) \psi_2 ( 0 ) }{2} 
\right\} {\cal S}^z 
\:\:\:\: , 
\label{rno.1}
\eneq
\noindent
with $\psi_1 ( x ) , \psi_2 ( x )$ being one-dimensional, chiral fermionic fields. 
$J_\perp, J_z$ are, respectively, the transverse and the longitudinal Kondo 
coupling strengths. On defining the associated dimensionless running coupling strengths 
${\cal J}_\perp ( D ) = \frac{a J_\perp}{v}$ and ${\cal J}_z ( D ) = \frac{a J_z }{v}$, the derivation of the 
RG equations for those running coupling has a long story and goes back to the 
original works on the subject \cite{anderson_yuval,ayh,Anderson_1970,hewson}. Specifically, on 
varying the energy cutoff $D$, one obtains that the corresponding variation of the running 
couplings is determined by the equations 

\begin{eqnarray}
 \frac{d {\cal J}_\perp ( D ) }{d \ln \left( \frac{D_0}{D} \right)} &=& {\cal J}_\perp ( D ) {\cal J}_z ( D ) \; ,  \nonumber \\
 \frac{d {\cal J}_z ( D ) }{d \ln \left( \frac{D_0}{D} \right)} &=& {\cal J}_\perp^2 ( D ) 
 \:  . 
 \label{rno.2}
\end{eqnarray}
\noindent
 The system in Eqs.(\ref{rno.2}) corresponds to the set of the standard  Kosterlitz-Thouless RG
  equations. To solve it,  one defines   the RG  invariant $H  = - {\cal J}_{z}^2 ( D ) + {\cal J}_{\perp}^2 ( D )$.
In particular, for $H=0$, one recovers the standard poor man's result for the 
RG equations in the case of isotropic Kondo effect \cite{Anderson_1970,hewson}.
In this case, one readily finds the solution in the form 

\beq
 {\cal J}_\perp ( D ) =  {\cal J}_\parallel ( D )  = \frac{{\cal J}_\perp ( D_0 ) }{1 -{\cal J}_\perp ( D_0 )  \ln \left( \frac{D_0}{D} \right)  }
\:\:\:\: , 
\label{rno0}
\eneq
\noindent
with the corresponding Kondo scale $D_K (=k T_K)$ given by 

\beq
D_K \sim D_0 e^{ - \frac{1}{{\cal J}_\perp ( D_0 ) } }
\:\:\:\: . 
\label{rno1}
\eneq
\noindent
At a generic value of $H$, as we show in Fig.\ref{region}, there are 
three relevant regions in the half-plane ${\cal J}_{ z } , {\cal J}_{\perp} > 0$. 
In detail, we have the following:

\begin{itemize}
 \item {\bf Region I:} this is defined for ${\cal J}_{  \parallel} ( D_0 ) > 0$ and 
 $H  < 0$.
 In this case, the integrated RG equations yield 
 
\begin{eqnarray}
 {\cal J}_{ z} ( D ) &=& \sqrt{ | H | } \: \left\{ \frac{  {\cal J}_{  z} ( D_0 ) + \sqrt{| H | }     
 + (  {\cal J}_{  z} ( D_0 ) - \sqrt{| H  | })  \left( \frac{D_0}{D} \right)^{ 2 \sqrt{ | H  |} }  }{ {\cal J}_{ z} ( D_0 ) 
 + \sqrt{| H  | } - (  {\cal J}_{  z} ( D_0 ) - \sqrt{| H | })  \left( \frac{D_0}{D} \right)^{ 2 \sqrt{ | H |}  }} \right\}
 \nonumber \\
  {\cal J}_{ \perp } ( D ) &=& 2 \sqrt{ | H  | } \: \left\{ \frac{ \sqrt{  {\cal J}^2_{  z} ( D_0 ) - | H | }     
 \left( \frac{D_0}{D} \right)^{  \sqrt{ | H |} }  }{ {\cal J}_{ z} ( D_0 ) 
 + \sqrt{|  H | } - (  {\cal J}_{  z} ( D_0 ) - \sqrt{| H | })  \left( \frac{D_0}{D} \right)^{ 2 \sqrt{ | H|}  }} \right\}
 \:\:\:\: . 
 \label{kt.a6}
\end{eqnarray}
 \noindent
Both running couplings increase as $D_0 / D$ gets large. Eventually, they hit a diverging point at the 
scale $D = D_{\rm KT}^{(1)}$, with 

\beq
D_{\rm KT}^{(1)} = D_0 \left\{ \frac{  {\cal J}_{  z} ( D_0 ) - \sqrt{| H | }}{  {\cal J}_{  z} ( D_0 ) + \sqrt{| H | }}
\right\}^\frac{1}{2 \sqrt{ |H  | }}
\:\:\:\: . 
\label{kt.a7}
\eneq
\noindent

\item {\bf Region II}: this is defined for $H  > 0$. In this case, one obtains 

\begin{eqnarray}
   {\cal J}_{  z} ( D ) &=& \sqrt{H } \: \tan \left\{ {\rm atan} \left( \frac{ {\cal J}_{  z} ( D_0 ) }{\sqrt{H}}
  \right) + \sqrt{H } \ln \left( \frac{D_0}{D} \right) \right\} \nonumber \\
   {\cal J}_{  \perp} ( D ) &=& \frac{ \sqrt{H } }{ \cos \left\{ {\rm atan} \left( \frac{ {\cal J}_{  z} ( D_0 ) }{\sqrt{H }}
  \right) + \sqrt{H } \ln \left( \frac{D_0}{D} \right) \right\} }
  \:\:\:\: . 
  \label{kt.a8}
\end{eqnarray}
\noindent
In the case $G_{  z} (D_0 ) < 0$, 
Eqs.(\ref{kt.a8}) imply a crossing of the vertical axis at the scale $D_{\rm cross}$ defined as 

\beq
D_{\rm cross} = D_0 \: \exp \left[ - \frac{1}{\sqrt{H}} \left|  {\rm atan} \left( \frac{ {\cal J}_{ z} ( D_0 ) }{\sqrt{H }}
  \right) \right| \right] 
\:\:\:\: . 
\label{kt.a9}
\eneq
\noindent
Both couplings diverge at the scale  $D = D_{\rm KT}^{(2)}$, with 

\beq
D_{\rm KT}^{(2)} = D_0 \: \exp \left[ - \frac{1}{\sqrt{H} } \left( \frac{\pi}{2} -
{\rm atan} \left( \frac{ {\cal J}_{  z} ( D_0 ) }{\sqrt{H}}  \right)  \right) \right] 
\:\:\:\: . 
\label{kt.a10}
\eneq
\noindent

 \item {\bf Region III:} this is defined for $ {\cal J}_{  z} ( D_0 ) < 0$ and 
 $H < 0$. In this case, the integrated RG trajectory take the form 
 
 \begin{eqnarray}
     {\cal J}_{  z} ( D ) &=&  \sqrt{ | H | } \: \left\{ \frac{ |  {\cal J}_{  z} ( D_0 ) | -  \sqrt{| H | }     
 + ( |  {\cal J}_{  z} ( D_0 ) | +  \sqrt{| H   | })  \left( \frac{D_0}{D} \right)^{ 2 \sqrt{ | H |} }  }{|  {\cal J}_{  z} ( D_0 ) | 
 - \sqrt{|  H  | } - ( |  {\cal J}_{  z} ( D_0 ) | + 
 \sqrt{| H | })  \left( \frac{D_0}{D} \right)^{ 2 \sqrt{ |  H |}  }} \right\}
 \nonumber \\
  {\cal J}_{ \perp } ( D ) &=& 2 \sqrt{ | H | } \: \left\{ \frac{ \sqrt{  {\cal J}^2_{  z} ( D_0 ) - | H | }     
 \left( \frac{D_0}{D} \right)^{  \sqrt{ | H  |} }  }{-|  {\cal J}_{ z} ( D_0 )| 
 + \sqrt{| H  | } + ( |  {\cal J}_{  z} ( D_0 ) | +
 \sqrt{| H  | })  \left( \frac{D_0}{D} \right)^{ 2 \sqrt{ |  H  |}  }} \right\}
 \:\:\:\: . 
 \label{kt.a11}
 \end{eqnarray}
\noindent
In this case, we see that the flow is no more towards a point at $\infty$, but we rather get 

\beq
\lim_{ \frac{D_0}{D} \to \infty } \: \left[ \begin{array}{c}
 {\cal J}_{  z} ( D ) \\  {\cal J}_{  \perp} ( D )                                        
                                      \end{array} \right] = 
\left[ \begin{array}{c}
- \sqrt{ | H | } \\ 0         
       \end{array} \right]
\:\:\:\:  .
\label{kt.a12}
\eneq
\noindent
\end{itemize}
From the analysis we perform above, it is natural to associate the onset of the 
Kondo regime (and the corresponding emergence  of a dynamically generated 
energy scale) to regions I and II, while region III is characterized by 
a flow toward  a manifold of ``trivial'' fixed points, continuously parametrized by 
$H$. In Fig.\ref{region} we provide a sketch of the RG 
 trajectories for $ {\cal J}_\perp ( D )$ and $ {\cal J}_z ( D )$ by particularly evidencing 
how, in the ``Kondo regions'' I and II, both running couplings flow to $\infty$. While this 
result is important for building a description of the corresponding Kondo fixed point, we 
now briefly review what are the possible effects of the ``non-Kondo''  terms in 
Eqs.(\ref{ishkz.1}) and (\ref{ishkz.2}). 
\noindent
A first additional term potentially appearing in the Kondo boundary Hamiltonian is the one-body, 
local scattering potential, which may just marginally change the single-particle 
phase shifts at the impurity location and, correspondingly, slightly renormalize 
the ac  conductance tensor with the Kondo interaction turned off, without essentially affecting the 
Kondo physics \cite{hewson}. Also, in the CK regime, one may 
get a  term corresponding to an effective, local 
magnetic field along the $z$ direction, either coupled to the impurity spin, 
or to the electronic spin density at the impurity location (or both). In general, 
these terms are known not to substantially affect the Kondo physics as long as 
the applied field $B$ is much lower than the energy scale associated to the Kondo temperature
\cite{costi,otte,grt}. In fact, this  
is the assumption we make here, as the effective $B$ field is determined by 
the direct density-density interaction within the wires, which is expected to be
much lower than the other energy scales in the system. Therefore, throughout all this 
paper, we consistently neglect 
the corresponding contributions to  the boundary Hamiltonian effectively describing the 
impurity dynamics.

To conclude the discussion of this appendix, we now consider  $H_{\rm DL}$ in Eq.(\ref{pera.11x}). Aside from the  one-body local 
scattering potentials, the only nontrivial interaction term is the direct local
density-density coupling, $\propto \kappa$. To deal with it, we therefore 
introduce the corresponding running coupling ${\cal K} ( D ) = \left( \frac{D_0}{D} \right)^{-1}
\kappa$. The RG equation for ${\cal K} ( D )$ to leading order in the running coupling is 
therefore given by 

\beq
\frac{ d {\cal K} ( D ) }{d \ln \left( \frac{D_0}{D} \right)} = - {\cal K} ( D )
\:\:\:\: , 
\label{rno.3}
\eneq
\noindent
which shows that this term is irrelevant and that, on lowering the cutoff, the corresponding 
running coupling scales as $D / D_0$. Accordingly, its effects can be consistently accounted 
for within a standard perturbative analysis in the coupling itself, which is what we 
made, when e.g. discussing the effects of $H_{\rm DL}$ on the dc-conductance tensor of 
the system.

 \section{ac  conductance tensor: details about the derivation in the various regions of 
the phase diagram of the system}
\label{ac_cond}

In this appendix, we explicitly  compute 
the equilibrium  ac conductance tensor in the various phases of  our system. To do so, we resort to 
a perturbative expansion around the fixed point corresponding to each one of the phases. When required
by the onset of a nonperturbative regime triggered by Kondo  interaction, we pertinently 
complement our analysis within renormalization group framework. In doing so, we assume that ac  frequency 
$\omega$ is large enough, so to avoid the suppression of the conductance  determined by the emergence of the finite time scale
(``Korringa time'' $\tau_K$ \cite{moca}), characterizing  correlations between subsequent impurity spin flips.
Indeed,  at frequencies $\omega \leq \tau_K^{-1}$), the 
spin conductance across the impurity (which, in our case, corresponds to the 
interwire conductance) is suppressed to 0 \cite{kinder,moca,tanaka,juk}. Thus,   in the following derivation  we assume that the condition $\omega , k T_K \gg k T$ always holds, 
 so that, as discussed below $\tau_K$ can be safely neglected and the perturbative calculation scheme 
applies \cite{moca,tanaka}. 

To encompass the various regimes of our system addressed in the paper, in the following we separately 
discuss the calculation for the effective spin- and charge-Kondo models, as well as for  the 
DL limit (which, in view of the apparent similarity between the corresponding boundary perturbations 
at the impurity, is expected to describe equally well the system in the vicinity of the spin- and of 
the charge-Kondo fixed points).

\subsection{Calculation of the ac conductance tensor in the Kondo regime}
\label{calko}

According to the derivation of  Appendix  \ref{pointl},
we begin by defining the analogs of Eqs.(\ref{ppl.15}) in our case, that is:
 
\begin{itemize}

\item For the spin-Kondo effective model,

\begin{eqnarray}
I_{1,<} &=& e v \{ \psi_{e , 1}^\dagger ( 0 ) \psi_{o , 1} ( 0 ) + \psi_{o , 1}^\dagger ( 0 ) \psi_{e , 1 }( 0 ) \}  
-  \frac{  e  }{2}\: \frac{ d {\cal S}^z }{d t } \; ,  \nonumber \\ 
I_{2,<} &=&  e v \{ \psi_{e , 2}^\dagger ( 0 ) \psi_{o , 2} ( 0 ) + \psi_{o , 2}^\dagger ( 0 ) \psi_{e , 2 }( 0 ) \}  
+  \frac{  e  }{2}\: \frac{ d {\cal S}^z}{d t } \; ,  \nonumber \\
I_{1,>} &=&  -  e v \{ \psi_{e , 1}^\dagger ( 0 ) \psi_{o , 1} ( 0 ) + \psi_{o , 1}^\dagger ( 0 ) \psi_{e , 1 }( 0 ) \} 
- \frac{  e  }{2}\: \frac{ d {\cal S}^z}{d t }  \; ,  \nonumber \\  
I_{2,>} &=& - e v \{ \psi_{e , 2}^\dagger ( 0 ) \psi_{o , 2} ( 0 ) + \psi_{o , 2}^\dagger ( 0 ) \psi_{e , 2 }( 0 ) \} 
+ \frac{  e  }{2}\: \frac{ d {\cal S}^z}{d t } 
\:  . 
\label{acc.x17}
\end{eqnarray}
\noindent

\item For the charge-Kondo effective model

\begin{eqnarray}
I_{1,<} &=& e v \{ \psi_{e , 1}^\dagger ( 0 ) \psi_{o , 1} ( 0 ) + \psi_{o , 1}^\dagger ( 0 ) \psi_{e , 1 }( 0 ) \}  - 
\frac{  e  }{2}\: \frac{ d {\cal T}^z}{d t } \; ,  \nonumber \\ 
I_{2,<} &=&  e v \{ \psi_{e , 2}^\dagger ( 0 ) \psi_{o , 2} ( 0 ) + \psi_{o , 2}^\dagger ( 0 ) \psi_{e , 2 }( 0 ) \}  
-\frac{  e  }{2}\: \frac{ d {\cal T}^z}{d t }  \; , \nonumber \\
I_{1,>} &=&   -  e v \{ \psi_{e , 1}^\dagger ( 0 ) \psi_{o , 1} ( 0 ) + \psi_{o , 1}^\dagger ( 0 ) \psi_{e , 1 }( 0 ) \} 
-  \frac{  e  }{2}\: \frac{ d {\cal T}^z}{d t }  \; , \nonumber \\ 
I_{2,>} &=&  - e v \{ \psi_{e , 2}^\dagger ( 0 ) \psi_{o , 2} ( 0 ) + \psi_{o , 2}^\dagger ( 0 ) \psi_{e , 2 }( 0 ) \} 
-\frac{  e  }{2}\: \frac{ d {\cal T}^z}{d t }
\: . 
\label{acc.x18}
\end{eqnarray}
\noindent
\end{itemize} 
In the SK case, as it emerges from 
Eqs.(\ref{acc.x17}), the whole set of   current-current 
correlation functions is determined by the correlations of the three operators 
${\cal O}_{\rm 1;KS} ( t  ) , {\cal O}_{\rm 2;KS} ( t  ), 
{\cal O}_{\rm S;KS} ( t  )$, given by 

\begin{eqnarray}
 {\cal O}_{\rm 1; KS} ( t ) &=& e v \{ \psi_{e , 1}^\dagger ( - v t ) \psi_{o , 1} ( - v t  ) + 
 \psi_{o , 1}^\dagger ( - v t  ) \psi_{e , 1 }( - v t  ) \}  \nonumber \\
  {\cal O}_{\rm 2; KS} (t ) &=&   e v \{ \psi_{e , 2}^\dagger ( - v t ) \psi_{o , 2} ( - v t  ) + 
  \psi_{o , 2}^\dagger ( - v t  ) \psi_{e , 2 }( - v t  ) \}   \nonumber \\
 {\cal O}_{\rm S;KS} (t) &=& e \frac{ d {\cal S}^z ( t) }{d t }
 \:\:\:\: . 
 \label{itc.1}
\end{eqnarray}
\noindent
To simplify the following calculations, we note that, due to the fact that the spin-Kondo Hamiltonian  
 only depends on the $\psi_{e , \alpha}$-fields and due to the identity 

\beq
e \frac{d {\cal S}^z ( t ) }{d t} = 2 i e J_{S , \perp}  \{ \psi_{e , 1 }^\dagger (  - v t  )
\psi_{e , 2} ( - v t ) {\cal S}^- ( t ) - \psi_{e , 2}^\dagger ( - v t ) \psi_{e , 1 } ( - v t ) 
{\cal S}^+ ( t ) \} 
\:\:\:\: ,
\label{itc.2}
\eneq
\noindent
we readily obtain that 

\beq
 {\cal G}_{( a , {\rm S  );  KS} } ( t - t' ) = - i \theta ( t - t' ) \:  \langle [ {\cal O}_{a ; {\rm KS} } ( t ) , 
 {\cal O}_{\rm S;KS} ( t' ) ] \rangle = 0 
 \;\;\;\; , 
 \label{itc.3}
\eneq
\noindent
for $a = 1,2$. By means of analogous arguments, we also obtain 

\beq
{\cal G}_{(1,2) ; {\rm KS}} ( t - t' ) = - i \theta ( t - t' ) \:   \langle [ {\cal O}_{1 ; {\rm KS} } ( t ) , 
 {\cal O}_{\rm 2;KS} ( t' ) ] \rangle = 0 
 \;\;\;\; , 
 \label{itc.4}
\eneq
\noindent
as well [Note that Eqs.(\ref{itc.3}) and (\ref{itc.4}) only depend on the functional form of 
the operators involved in the calculations. Therefore, they are exact at any order of 
the perturbative expansion in $J_S$]. Therefore, we only need to compute 
``diagonal'' correlation functions. Retaining the leading perturbative contributions in 
$J_{S , \perp}$,  we stop our derivation to order $J_{S , \perp}^2$. We therefore obtain 

\begin{eqnarray}
 {\cal G}_{\rm  S ; KS} ( t - t' ) &=& - i e^2 \theta ( t - t' ) \: \left\langle \left[  \frac{d {\cal S}^z ( t ) }{d t } , 
 \frac{d {\cal S}^z ( t' ) }{d t' } \right] \right\rangle    \nonumber \\
 &=& - \frac{4 i e^2 J_{S , \perp}^2 }{ ( 2 \ell )^2} \: \theta ( t - t' ) \: \sum_{\epsilon  , \epsilon ' } \: f ( \epsilon  ) f ( \epsilon' ) \: \{ 
 e^{ i  ( \epsilon + \epsilon'   ) ( t - t' ) } -  e^{ - i  ( \epsilon + \epsilon' ) ( t - t' ) } \}
 \:\:\:\: , 
 \label{itc.5}
\end{eqnarray}
\noindent
with $f ( \epsilon )$ being Fermi distribution function.
When resorting to Fourier space, Eq.(\ref{itc.5}) implies

\beq
 {\cal G}_{\rm  S ; KS} ( \omega ) = \frac{4 e^2 J_{S , \perp}^2}{ ( 2 \ell )^2} \: \sum_{\epsilon , \epsilon'} \: f ( \epsilon ) f ( \epsilon' ) \: 
 \left[ \frac{1}{\omega +  ( \epsilon + \epsilon' ) - i \eta} - \frac{1}{\omega -  ( \epsilon + \epsilon' ) - i \eta} \right] 
 \:\:\:\: , 
 \label{itc.6}
 \eneq
 \noindent
 with $\eta = 0^+$.  Similarly, one obtains 
 
\begin{eqnarray}
&& {\cal G}_{(a,a) ; {\rm KS}} (\omega) = - i \int \: d (t-t') \: e^{ i \omega ( t-t' )} \: 
 \theta ( t - t' ) \:   \langle [ {\cal O}_{a ; {\rm KS} } ( t ) , 
 {\cal O}_{a; \rm  KS} ( t' ) ] \rangle  \nonumber \\
 && = - \frac{i e^2 \omega }{2 \pi} \: \left\{ 1 - \frac{3 J_{S , \perp}^2}{ v^2} \right\}
 \:\:\:\: . 
 \label{iii.1}
\end{eqnarray}
\noindent
Equations (\ref{itc.6}) and (\ref{iii.1}) are all we need to perturbatively compute the ac conductance. 
As a result, we obtain

\begin{eqnarray}
 {\bf G}_{(1,>);(1,<)} ( \omega ) &=& - \frac{e^2}{2 \pi } \: \left\{ 1 -\frac{  J_{S , \perp}^2}{  v^2} \right\} \; , \nonumber \\
 {\bf G}_{(2,>);(1,<)} ( \omega ) &=&    - \frac{e^2}{ 2 \pi}  \frac{  J_{S , \perp}^2}{  v^2} 
 \:  . 
 \label{itc.9}
\end{eqnarray}
\noindent
Similarly, in the charge Kondo case one obtains

\begin{eqnarray}
 {\bf G}_{(1,>);(1,<)} ( \omega ) &=& - \frac{e^2}{2 \pi  } \: \left\{ 1 - \frac{  J_{C , \perp}^2}{v^2} \right\} \; ,  \nonumber \\
 {\bf G}_{(2,>);(1,<)} ( \omega ) &=&     \frac{e^2}{ 2 \pi}  \frac{  J_{C , \perp}^2}{v^2} 
 \:  . 
 \label{itc.10}
\end{eqnarray}
\noindent
Equations (\ref{itc.9}) and (\ref{itc.10}) are expected to apply in the frequency range 
$\omega   \gg k T_K$. As $\omega$ goes down toward $k T_K$, one expects that the net effect of the Kondo 
interaction is to renormalize the coupling strengths $J_{S , \perp} , J_{C , \perp}$ as discussed in 
Appendix \ref{KT}, with  $\omega$ 
corresponding to the running scale $D$. Taking this into account, 
Eqs.(\ref{itc.9}) and (\ref{itc.10}) become

\begin{eqnarray}
 {\bf G}_{(1,>);(1,<)} ( \omega  ) &=& - \frac{e^2}{2 \pi } \: \left\{ 1 - {\cal J}_{S , \perp}^2 ( \omega )  \right\} \; ,  \nonumber \\
 {\bf G}_{(2,>);(1,<)} ( \omega ) &=&  - \frac{e^2}{2 \pi } {\cal J}_{S , \perp}^2 ( \omega )
 \:  ,
 \label{itc.11}
\end{eqnarray}
\noindent
in the spin Kondo case, and 

\begin{eqnarray}
 {\bf G}_{(1,>);(1,<)} ( \omega ) &=& - \frac{e^2}{2 \pi } \: \left\{ 1 -
 {\cal J}_{C , \perp}^2 ( \omega ) \right\} \; , \nonumber \\
 {\bf G}_{(2,>);(1,<)} ( \omega ) &=&   \frac{e^2}{2 \pi }  {\cal J}_{C , \perp}^2 ( \omega ) 
 \:  , 
 \label{itc.12}
\end{eqnarray}
\noindent
in the charge Kondo case, with 
${\cal J}_{S (C) , \perp (z) } = \frac{a J_{S (C) , \perp (z)}}{v}$,  
 $H_{S (C)}   = - {\cal J}_{S (C)  , z}^2  + {\cal J}_{S (C) , \perp}^2$,  
 
\beq
{\cal J}_{S (C)  , \perp}  (  \omega  ) = \frac{2 \sqrt{ | H_{S (C)}  | } \left( \frac{k T^{S (C)}_K}{\omega} \right)^{\sqrt{| H_{S (C)} |} }  }{1 - \left( \frac{k T^{S(C)}_K}{\omega}
\right)^{2 \sqrt{| H_{S (C)} |}} } 
\:\:\:\: , 
\label{csc.2y}
\eneq
\noindent
 and the Kondo temperature $k T_K^{S (C)}  =   D_0     
  \left\{ \frac{  {\cal J}_{S (C) , z  } ( D_0 ) - \sqrt{| H_{S (C)}  | } }{  {\cal J}_{ S (C) , z } ( D_0 ) + \sqrt{| H_{S (C)} | }} \right\}$. 

Equations (\ref{itc.11}) and (\ref{itc.12}) provide the perturbative result for the ac conductances of interest for 
our work. 
Thus, we conclude that the only relevant effect of 
the Kondo dynamics on the interlead ac conduction properties of the system is encoded in the 
dynamical impurity spin Green's functions ${\cal G}_{\rm S ; KS} ( t - t' )$ and 
${\cal G}_{\rm S ; KC}  ( t - t' )$. Indeed,  typically, in 
problems dealing with ac transport across a  Kondo-type  impurity, the ac conductance 
is directly related to the (time derivatives of the) impurity spin Green's functions as 
discussed, for instance, in Ref.[\onlinecite{moscio}]  in the context of microwave 
scattering at a quantum impurity in a Josephson junction array.

\subsection{Perturbative calculation of the conductance in the disconnected lead limit}
\label{perdis}

In the disconnected lead limit, the leading boundary perturbation in the lattice framework is given by 
$H_{\rm DL}$ in Eq.(\ref{reno0.2}), which we adopt in the following as our perturbing boundary 
Hamiltonian, with the symmetric simplification $\lambda_1 = \lambda_2 = \lambda$ and with 
(in terms of the lattice model Hamiltonian parameters)

\begin{eqnarray}
 \lambda &\sim&   \frac{t^2}{2 \delta} \; ,   \nonumber \\
 \kappa &\sim& \left( \frac{  t }{ 2 \delta } \right)^4 \: \delta 
 \:. 
 \label{perdi.2}
\end{eqnarray}
\noindent
(See Appendix \ref{pertdelta} for details). $H_{\rm DL}$ commutes with the total 
charge density at $x=0$ in both leads. Therefore, the current operators defined in 
  Eqs.(\ref{ppl.15}) are now  given by 

\begin{eqnarray}
I_{1,<} &=& e v \{ \psi_{e , 1}^\dagger ( 0 ) \psi_{o , 1} ( 0 ) + \psi_{o , 1}^\dagger ( 0 ) \psi_{e , 1 }( 0 ) \}  \; , 
\nonumber \\ 
I_{2,<} &=&  e v \{ \psi_{e , 2}^\dagger ( 0 ) \psi_{o , 2} ( 0 ) + \psi_{o , 2}^\dagger ( 0 ) \psi_{e , 2 }( 0 ) \} \; ,  
\nonumber \\
I_{1,>} &=&  -  e v \{ \psi_{e , 1}^\dagger ( 0 ) \psi_{o , 1} ( 0 ) + \psi_{o , 1}^\dagger ( 0 ) \psi_{e , 1 }( 0 ) \} \; , 
 \nonumber \\  
I_{2,>} &=& - e v \{ \psi_{e , 2}^\dagger ( 0 ) \psi_{o , 2} ( 0 ) + \psi_{o , 2}^\dagger ( 0 ) \psi_{e , 2 }( 0 ) \} 
\: . 
\label{perdi.3}
\end{eqnarray}
\noindent
A first important observation arising at a first glance to Eqs.(\ref{perdi.2}) and (\ref{perdi.3}) is 
that, since $H_{\rm DL}$ only contains the $-e$-fields at $x=0$ while $I_{a,>}$ and 
$I_{a,<}$ are linear in terms of $\psi_{o,a} ( 0 )$ and $\psi_{o,a}^\dagger ( 0 )$, one 
obtains the exact result that ${\bf G}_{(2,>);(1,<)} ( \omega ) = {\bf G}_{(1,>);(2,<)} ( \omega ) = 0$, 
at any order in $\kappa$. As for what concerns corrections to the ac conductances diagonal in 
the lead index, following the analysis of Appendix \ref{KT}, we expect a perturbative calculation in $\kappa$ 
to provide reliable results for the ac conductance tensor. Setting $\kappa = 0$, our system 
traces back to the one we discuss in Appendix \ref{pointl}, with $V_{od} = 0$, 
$V_d =  \frac{t^2}{ \delta} $. Therefore, setting  $\kappa=0$, we obtain 
the conductances ${\bf G}^{(0)}_{(1,>);(1,<)} ( \omega )  $,
given by 

\beq
{\bf G}^{(0)}_{(1,>);(1,<)} ( \omega )  = - \frac{e^2}{2 \pi} \: \left[ \frac{  v^2 \delta^2}{  v^2 \delta^2 + t^4} \right]  \equiv - \frac{e^2}{2 \pi} \cos^2 ( \delta_\kappa)
\:\:\:\: , 
\label{perdi.4}
\eneq
\noindent 
with $\delta_\kappa$ single-fermion phase shift at the DL fixed point. 
To second-order in $\kappa$, Eq.(\ref{perdi.4}) is corrected into 

\beq
{\bf G}_{(1,>) ; (1,<)} ( \omega ) \approx -   \frac{e^2}{2 \pi} \:  
\left\{ \cos^2 ( \delta_\kappa) - \frac{8 \cos ( 2 \delta_\kappa)  \kappa^2 \omega^2}{3 \pi^2 v^2}   \right\}
\:\:\:\: , 
\label{ps.5}
\eneq
\noindent
 that is, the result we used in the main text.

\bibliography{iso_biblio}
\end{document}